\documentclass[preprint,12pt]{elsarticle}

\usepackage{amssymb}
\usepackage{amsmath}
\usepackage{graphicx}
\usepackage{subfig}
\usepackage{multirow}
\usepackage{threeparttable}
\usepackage{multicol}
\usepackage{array}
\usepackage{color}
\newcolumntype{M}[1]{>{\centering\arraybackslash}m{#1}}
\newcolumntype{L}[1]{>{\raggedright\arraybackslash}m{#1}}

\journal{International Journal of Electrical Power \& Energy Systems}

\begin{document}

\begin{frontmatter}



\title{Online MPC-Based EMS for Microgrids With Multi-Modal PV Forecasting and Cone-Relaxed Power-Flow Constraints}


\author[1]{Hanyang He} 
\author[2]{John Harlim} 
\author[3]{Daning Huang}
\author[1]{Yan Li}

\affiliation[1]{organization={Department of Electrical  Engineering, Pennsylvania State University},
            city={State College},
            postcode={16803}, 
            state={PA},
            country={USA}}

\affiliation[2]{organization={Department of Meteorology and Atmospheric Science, Institute for Computational and Data Sciences, Pennsylvania State University},
            city={State College},
            postcode={16803}, 
            state={PA},
            country={USA}}

\affiliation[3]{organization={Department of Aerospace Engineering, Pennsylvania State University},
            city={State College},
            postcode={16803}, 
            state={PA},
            country={USA}}
            
\begin{abstract}
The large-scale integration of photovoltaic (PV) generation, whose output is highly sensitive to weather conditions and climate variability, poses significant challenges to the secure, economic, and stable operation of microgrids. To enhance the operational resilience of microgrids against significant deviations between actual PV generation and day-ahead forecasts due to weather uncertainty, this paper proposes an online model predictive control (MPC)-based energy management system (EMS) capable of handling PV-forecasting deviations and associated power-flow constraint violations.
Within the proposed framework, the MPC is equipped with a multi-modal dictionary-guided anisotropic kernel ridge regression (MMDG-AKRR) forecasting model that improves the generalizability of PV power forecasting across heterogeneous weather conditions. By leveraging online similarity-driven model weighting, the proposed forecasting approach enables timely correction of PV power trajectories when significant deviations from day-ahead predictions occur. Meanwhile, a second-order cone relaxation of power-flow constraints, explicitly formulated in terms of bus-voltage and branch-current magnitudes, is embedded into the MPC optimization to ensure grid-secure and computationally efficient dispatch decisions.
Comprehensive comparative studies demonstrate that, relative to conventional neural-network-based and kernel-based forecasting methods, the proposed MMDG-AKRR model achieves better short-horizon prediction adaptability and enhanced robustness to PV forecasting uncertainties. Moreover, the proposed MMDG-AKRR-assisted MPC-EMS framework delivers a more favorable balance between operational security and economic performance than traditional day-ahead EMS schemes and linear MPC-based approaches. Compared with nonlinear MPC formulations that explicitly enforce nonlinear power-flow constraints, the proposed framework substantially reduces computational burden while maintaining high-quality solutions, highlighting its practical applicability for real-time microgrid energy management.
\end{abstract}

\begin{graphicalabstract}
\includegraphics[width=\textwidth]{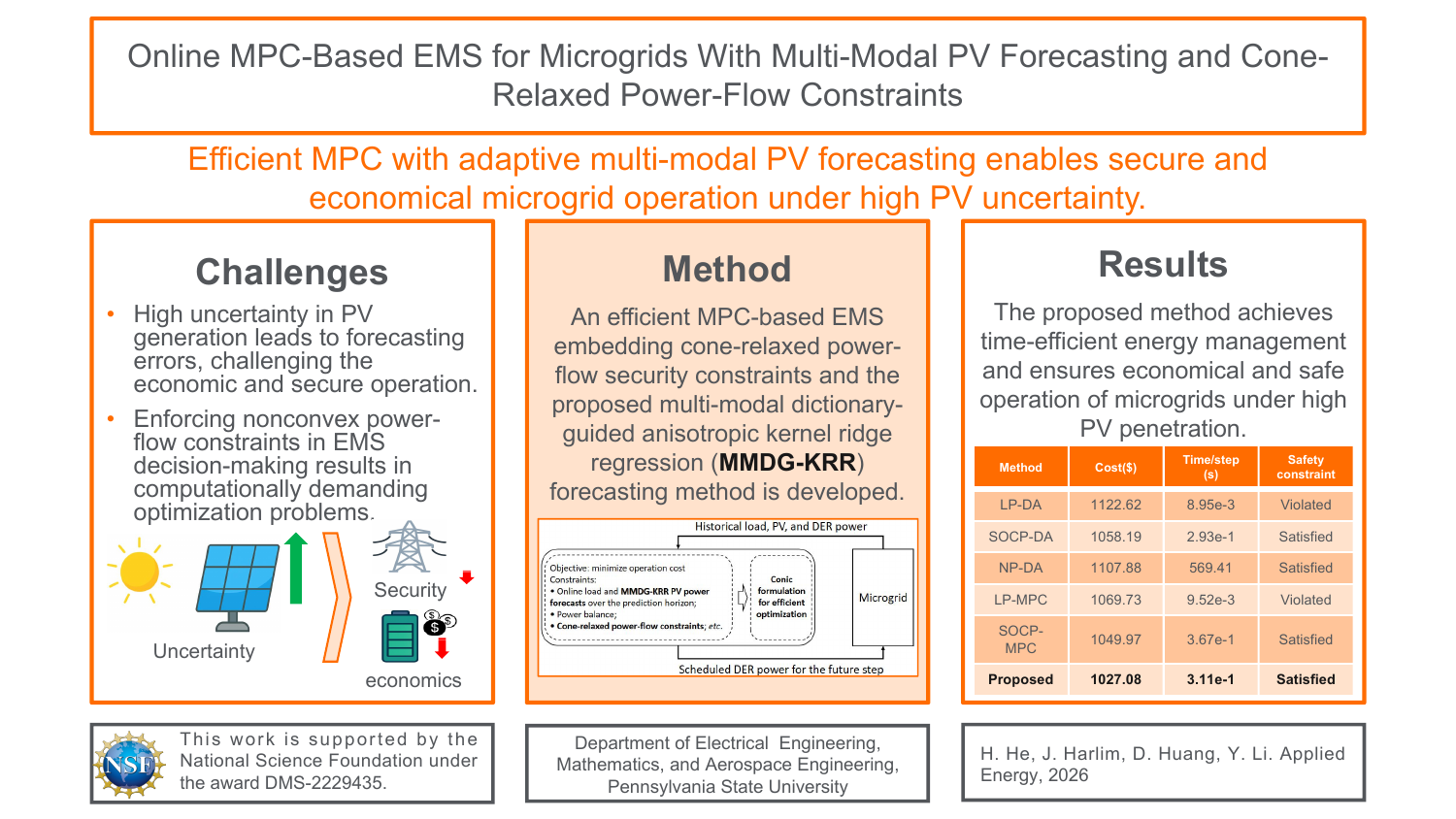}
\end{graphicalabstract}

\begin{highlights}
\item A multi-modal dictionary-guided anisotropic kernel ridge regression (MMDG-AKRR) forecasting method is developed to enhance online adaptability under highly uncertain photovoltaic power trajectories.
\item A computationally efficient online MPC-EMS is constructed by embedding an SOCP-relaxed branch-flow security layer into the rolling optimization.
\item Numerical studies on 10-, 18-, and 33-bus microgrids demonstrate improved forecasting robustness, enhanced operational security, and reduced computational cost compared with conventional EMS schemes.
\end{highlights}


\end{frontmatter}


\section*{Nomenclature}

\begin{description}
  \item[$C_{\text{buy}}$] \hspace{0.2em}Electricity purchase price.
  \item[$C_{\text{deg}}$] \hspace{0.2em}Battery degradation cost.
  \item[$C_{\text{DG}}$] \hspace{0.2em}Diesel generator cost.
  \item[$C_{\text{sell}}$] \hspace{0.2em}Electricity selling price.
  \item[$E_{\text{max,}i}$] \hspace{0.2em}Battery capacity of the $i$th storage.
  \item[$\eta_{\text{batt}}^{\text{ch/dis}}$] \hspace{0.2em}Charging/discharging efficiency of storage.
  \item[$N_{\mathrm{batt}}$] \hspace{0.2em}The number of the battery.
  \item[$N_{\mathrm{br}}$] \hspace{0.2em}The number of the branches.
  \item[$N_{\mathrm{tp}}$] \hspace{0.2em}The number of load types.
  \item[$\Omega_{\text{batt}}$] \hspace{0.2em}Set of battery buses.
  \item[$\Omega_{\text{br/bus}}$] \hspace{0.2em}Set of branches/buses.
  \item[$\Omega_{\mathrm{tp}}$] \hspace{0.2em}Set of load types.
  \item[$P_{\text{buy/sell}}$] \hspace{0.2em}Power purchased from/sold to the main grid.
  \item[$P_{\text{slk}}$] \hspace{0.2em}Power exchanged with the main grid.
  \item[$P_{\text{dis/ch,}i}$] \hspace{0.2em}Discharging/charging power of the $i$th storage.
  \item[$P_{\text{batt,}i}$] \hspace{0.2em}Net power exchanged between the $i$th energy storage unit and the grid.
  \item[$P_{\text{DG,}i}$] \hspace{0.2em}Power of the $i$th diesel generator.
  \item[$P_{\text{load,}i}$] \hspace{0.2em}Predicted load power of the $i$th load with.
  \item[$P_{\text{res,}i}$] \hspace{0.2em}Active power of the $i$th renewable source.
  \item[$\sigma_{\text{SoC,}i}$] \hspace{0.2em}State of charge (SoC) of the $i$th storage. 
  \item[$\tilde{l}$] \hspace{0.2em}Auxiliary variable for squared branch current $|I|^2$.
  \item[$\tilde{v}$] \hspace{0.2em}Auxiliary variable for squared node voltage $|V|^2$.
\end{description}

\section{Introduction}

The large-scale integration of photovoltaic (PV) generation into microgrids enhances system flexibility, reduces carbon emissions, and improves overall economic performance~\cite{Parhizi2015,RodriguezGil2024EMS}. At the same time, the strong climate sensitivity of PV generation introduces significant uncertainty into system operation. In particular, PV output is highly affected by weather conditions such as cloud coverage, which substantially increases the difficulty of accurate forecasting and calls for energy management systems (EMSs) with strong adaptability to PV power fluctuations. Moreover, the high penetration of renewable energy sources (RESs) with variable output intensifies the operational security challenges of microgrids, especially for networks with non-negligible electrical distances. Under such conditions, the EMS is required to explicitly account for power-flow security constraints to ensure safe and reliable operation.

Addressing the first challenge requires the selection of an appropriate EMS architecture. 
In general, EMS strategies can be categorized into two complementary paradigms: day-ahead planning and online management. Day-ahead EMS focuses on medium- to long-term scheduling of power generation, load profiles, and electricity prices based on forecasted information, with the primary objective of minimizing overall system operating costs~\cite{Duc2024,PowerSystemMMG2018,HARSH2021101225}. Owing to its relatively long decision intervals, day-ahead planning typically imposes modest computational requirements and is well-suited for large-scale power systems with complex network models. As a planning-oriented strategy, it plays a critical role in establishing an economically efficient and operationally feasible baseline, particularly for systems with reasonably predictable operating conditions.
However, real-time disturbances and forecast uncertainties may lead to deviations from the planned schedules, especially in microgrids with high penetration of RESs. This motivates the integration of online EMS to complement day-ahead planning by enabling timely corrective actions that enhance both operational reliability and economic performance.

Online EMS approaches can be further categorized into two main types: rule-based methods and optimization-based methods.
Rule-based EMSs typically rely on predefined heuristics and operating rules derived from prior experience or engineering knowledge, which makes them simple to implement and computationally efficient.
For example, a single-step rule-based strategy was introduced in~\cite{Meiqin2014} for online scheduling. A decision tree-based EMS method was proposed in~\cite{Gunaid2016}. A rule-based controller incorporating multi-step wind power forecasts was developed in \cite{Guo2016TSG} to manage stand-alone microgrids. An adaptive, data-driven, rule-based EMS was proposed in \cite{Venayagamoorthy2016} by using a reinforcement learning framework. 
However, their performance strongly depends on the quality of the predefined rules, and they generally lack the flexibility to handle complex operating conditions, multiple conflicting objectives, and rapidly changing system states. These limitations motivate the adoption of optimization-based EMS frameworks.

Model predictive control (MPC) is the most suitable optimization-based EMS approach for this study due to its receding-horizon structure and its ability to explicitly handle multiple objectives and operational constraints~\cite{EndtoEnd_CasagrandeTCST2025,HU2021110422,Hu2024}. 
Numerous studies have demonstrated the effectiveness of MPC-based EMSs in coordinating distributed energy resources under dynamic operating conditions. 
For instance, the work in~\cite{Hu2024} comprehensively reviewed MPC applications in photovoltaic-storage-load scheduling, realizing cost reduction and battery life extension. An MPC-based EMS for vehicle-to-grid charging stations was proposed in~\cite{Ponce2024}, enabling seamless transition between grid-connected and islanded modes and reducing the cost. More importantly, online forecast updates within MPC enable adaptive responses to unusual weather conditions, thereby mitigating renewable variability and reducing operating costs~\cite{Mirletz2023ForecastErrors}.
However, the performance of MPC critically depends on the accuracy of power forecasting models, which motivates the development of suitable time-series learning methods to support reliable and adaptive online decision-making.


A wide range of time-series learning methods has been developed for renewable-power and load forecasting. From the forecasting-structure perspective, these methods can be broadly categorized into sequence-to-sequence and autoregressive sequence-to-point paradigms. {Sequence-to-sequence methods directly map a historical input sequence to a future prediction sequence and are widely adopted in high-capacity neural-network models, such as CNN-, LSTM-, GRU-, and other attention-based architectures~\cite{Deep_Ali2023,ResidentialCNN_Marcus2018,Israni2025LSTMGRU,Zhou2021Informer}. These models can capture complex temporal dependencies and implicit temporal patterns when sufficient training data are available. However, their modal adaptation is usually learned inside a single high-capacity end-to-end model, and the learned modes are not explicitly organized as representative PV operating profiles. For the online EMS problem considered in this paper, the available profile-level PV training data are limited, and the test profiles may deviate significantly from the representative training distribution due to abrupt weather changes. Under such data-scarce and distribution-shifted conditions, large black-box sequence models may suffer from unstable extrapolation and overfitting, especially when the online controller must repeatedly update forecasts from short measured input windows.}

In contrast, autoregressive sequence-to-point methods perform rolling one-step-ahead prediction using shorter input sequences, making them more suitable for cold-start prediction, limited historical records, {and frequent online updates in MPC.} Representative approaches include neural-network-based models~\cite{Filik2007_STLF_AR_ANN,GNN_di2023,STCN_Siya2025} and kernel-based models~\cite{MKRR_di2020,KRLS_Chen2010}. Neural-network-based methods provide flexible nonlinear approximation but usually require more data to maintain robust generalization. Kernel-based methods, particularly KRR-type models, are more data-efficient and stable in small-sample settings, { but conventional isotropic or single-distribution kernel models have limited ability to distinguish heterogeneous feature sensitivities and adapt to multiple PV operating modes. Therefore, a forecasting model that combines sample-efficient kernel learning with explicit modal adaptation is needed for online MPC-based EMS under PV forecast deviations.}

More critically, although existing forecasting methods can achieve satisfactory performance when the training and testing profiles follow similar distributions, they remain inadequate for PV trajectories characterized by pronounced randomness and multiple operating modes. A single forecasting model trained around a representative profile may exhibit limited extrapolation capability when confronted with atypical or low-generation PV profiles. This motivates the proposed MMDG-AKRR model, which uses an anisotropic kernel to distinguish the power-sequence and time-stamp features, and further organizes multiple AKRR base models into a modal dictionary. {Different from attention-based sequence models that learn latent temporal patterns within one global model, the proposed MMDG-AKRR explicitly embeds several base models trained for representative PV operating profiles into a dictionary. During online MPC operation, the current measured input sequence is compared with the representative modal profiles, and similarity-driven weights are used to combine these local base models. In this way, the proposed method serves as a data-efficient and interpretable online correction mechanism for multi-modal PV trajectories, rather than a large end-to-end deep sequence predictor.}

The second challenge, mentioned in the first paragraph, lies in developing an MPC-based EMS framework that explicitly accounts for power-flow security constraints while maintaining a favorable balance between computational efficiency and solution quality. For microgrids with short electrical distances, network losses and voltage drops are typically negligible. In such cases, neglecting the nonlinear power-flow model preserves the linear structure of the MPC optimization problem, thereby ensuring high computational efficiency and reliable solution quality. As a result, many existing MPC-based EMS approaches simplify the network power-flow constraints to a node-injection power balance relationship~\cite{Chengquan_TSG2018,Nair2020,Valencia2016,Zhuoli2022,Xie2021,EVbuildingEMSMPC_TSG2022,DPMarketEMSMPC_TSG2021}.

However, for medium- to large-scale microgrids with non-negligible electrical distances, such as campus microgrids, the high penetration of PVs necessitates the explicit consideration of network security constraints to ensure safe operation. Directly embedding nonlinear power-flow equations into the MPC formulation results in a nonconvex optimization problem, which significantly degrades computational tractability and solution quality. To address this issue, a second-order cone programming (SOCP) relaxation technique was proposed in~\cite{Farivar2013IEEE_TPS}, providing a convex formulation that enables the incorporation of power-flow constraints while preserving computational efficiency. Nevertheless, this technique has primarily been applied to day-ahead EMS formulations~\cite{QianXunTSE2025} that are insensitive to real-time computational requirements, and its integration into online MPC-based EMS frameworks remains largely unexplored.

To address the above challenges, this paper proposes a power-flow-constrained online MPC-based EMS framework integrated with the proposed MMDG-AKRR forecasting method. The main contributions are summarized as follows:
\begin{itemize}
\item \textbf{{A multi-modal dictionary-guided AKRR forecasting method is developed for online PV trajectory correction.}} {The proposed MMDG-AKRR combines an anisotropic kernel for heterogeneous power-sequence and time-stamp features, a dictionary of AKRR base models trained under representative PV operating modes, and an online similarity-driven weighting strategy based on the measured initial PV sequence. Different from large end-to-end neural sequence predictors, the proposed model is designed as a data-efficient and interpretable correction mechanism for PV profiles with limited training samples and significant weather-driven distribution shifts.}

\item \textbf{{A forecasting-aware SOCP-constrained online MPC-EMS framework is constructed.}} {The proposed EMS couples online MMDG-AKRR forecast updates with rolling-horizon battery/grid dispatch, while embedding an SOCP-relaxed branch-flow security layer to account for voltage and branch-current constraints. This formulation enables repeated online optimization with improved PV-adaptation capability and tractable computational cost.}

\item \textbf{{A controlled comparative validation is conducted to clarify the roles of the forecasting and EMS components.}} {The forecasting module is compared with NN, MMDG-NN, MKRR, and AKRR baselines to evaluate the effects of the modal dictionary and anisotropic kernel. The EMS framework is compared with LP-based, SOCP-based, and nonlinear power-flow-constrained day-ahead/MPC schemes on 10-, 18-, and 33-bus microgrids to assess forecasting adaptability, grid-security performance, operating cost, and computational efficiency.}
\end{itemize}

{ The novelty of this work should therefore be interpreted at two levels. The main methodological novelty lies in the MMDG-AKRR forecasting model, which provides a data-efficient multi-modal PV correction mechanism for online MPC. The EMS-side contribution does not claim MPC or SOCP branch-flow relaxation as standalone new techniques. Instead, it lies in integrating the proposed online forecasting correction with an SOCP-constrained rolling MPC workflow for grid-secure microgrid energy management. This integrated design aims to balance PV-forecasting adaptability, operational security, economic performance, and real-time computational tractability. }

The remainder of this paper is organized as follows. Section II formulates the overall problem framework. The forecasting models used for MPC are presented in Section III. Section IV details the relaxed branch flow model and the conic formulation. Section V discusses the case study results, and Section VI concludes the paper.

\section{Problem Formulation}\label{sec_wholeframe}

This section first presents a general MPC framework. The proposed MMDG-AKRR-based MPC-EMS is then developed within this framework to provide a clear algorithmic structure.

\subsection{General Framework of MPC}

MPC is a receding horizon control strategy. A discrete framework of the MPC problem is given in \eqref{eq_MPCgeneral}, with prediction horizon $N_{\text{pre}}$, prediction step size $\Delta t_{\text{p}}$, and initial step $k\in\Omega_{\text{d}}$, where $\Omega_{\text{d}}=\{0,1,\dots,K_{\text{d}}-1\}$ is the decision step set with decision interval $\Delta t_{\text{d}}$.
\begin{subequations}\label{eq_MPCgeneral}
    \begin{align}
        \min_{\mathbf{u}^{[k\to N_{\text{end}}^{[k]}]}}J&=\sum_{n=k}^{k+N_{\text{pre}}-1}\ell(\mathbf{x}^{[n|k]},\mathbf{y}^{[n|k]},\mathbf{u}^{[n|k]})\\
        \text{s.t.: }\:\:\:\:\;&\;\notag\\
        \mathbf{x}^{[n+1|k]}&=\mathbf{x}^{[n|k]}+\mathbf{f}_{\text{Dyn}}(\mathbf{x}^{[n|k]},\mathbf{y}^{[n|k]},\mathbf{u}^{[n|k]};\Delta t_{\text{p}})\label{eq_fdyb}\\
        \mathbf{y}^{[n|k]}\;\;\;&=\mathbf{f}_{\text{Alg}}(\mathbf{x}^{[n|k]},\mathbf{y}^{[n|k]},\mathbf{u}^{[n|k]})\label{eq_falg}\\
        \;\mathbf{x}^{[n|k]}\in&\Omega_{\mathbf{x}},\;\mathbf{y}^{[n|k]}\in\Omega_{\mathbf{y}},\mathbf{u}^{[n|k]}\in\Omega_{\mathbf{u}},\label{eq_mpc_const}
    \end{align}
\end{subequations}
where the notation $\Box^{[n|k]}$ denotes the value of a variable at prediction step $n$, given the initial decision step $k$. $N_{\text{end}}^{[k]}=k+N_{\text{pre}}-1$ is the last index of $\mathbf{u}$. Equation \eqref{eq_fdyb} denotes the dynamic constraints $C_{\text{dyn}}$, including the state of charge (SoC) dynamics and power forecasting models in the EMS setting. Equation \eqref{eq_falg} represents algebraic constraints $C_{\text{alg}}$ such as power balance and grid security constraints. Equation \eqref{eq_mpc_const} collects other inequality constraints $C_{\text{ieq}}$ on state, output, and control variables.


\subsection{Overview of the Proposed MPC-based EMS}\label{sec_problemFormulation}
\begin{figure}
    \centering
    \includegraphics[width=0.7\linewidth]{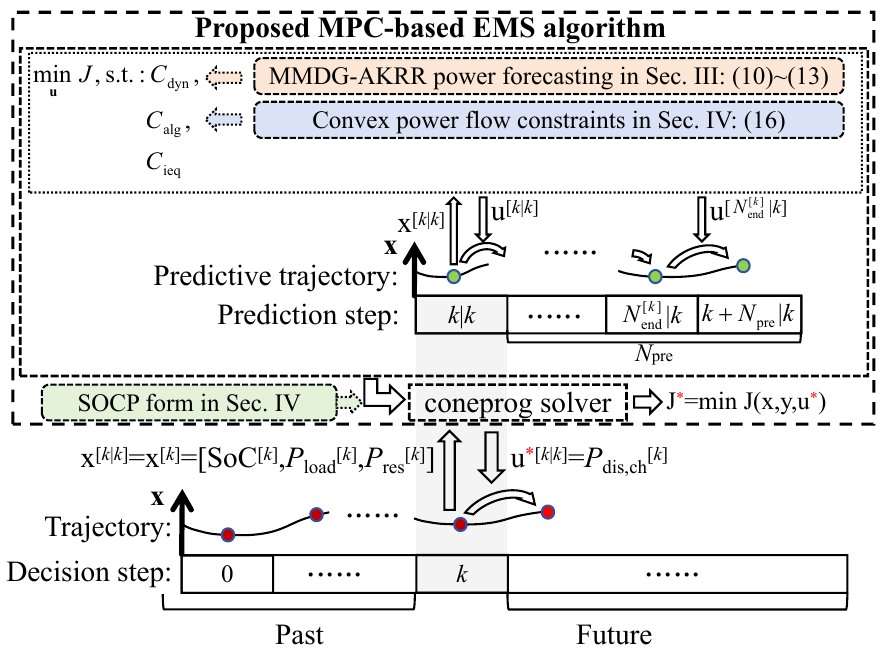}
    \caption{Overview of the MPC-based EMS algorithm.}
    \label{fig_Wholeframe}
\end{figure}

This section presents the detailed optimization model, as shown in Fig.~\ref{fig_Wholeframe}, for the proposed MPC-EMS based on the framework in \eqref{eq_MPCgeneral}. 
The decision variables, state variables, and algebraic variables are listed below, respectively,
\begin{align}
    \mathbf{u}=&\{{P}_{\text{buy}},{P}_{\text{sell}},
    {P}_{\text{dis,}i\in\Omega_{\text{batt}}},{P}_{\text{ch,}i\in\Omega_{\text{batt}}},\,{P}_{\text{DG,}i\in\Omega_{\text{DG}}},\notag\\
    &\tilde{v}_{i\in\Omega_{\text{bus}}},\tilde{l}_{ij\in\Omega_{\mathrm{br}}},\,P_{ij\in\Omega_{\text{br}}},\Delta C_{\text{inc,}p\in\{1,\dots,N_{\mathrm{tp}}\}}\},\notag\\
    \mathbf{x}=&\{\sigma_{\text{SoC,}i\in\Omega_{\mathrm{batt}}},{P}_{\text{res,}j\in\Omega_{\mathrm{res}}},{P}_{\text{load,}j\in\Omega_{\mathrm{load}}}\},\notag\\
    \mathbf{y}=&\{P_{j\in\Omega_{\mathrm{bus}}}\},
\end{align}
The running cost $\ell$ is
\begin{align}
    \ell&=
\Delta t_{\text{d}} \Big(
- C_{\text{sell}}^{[n|k]} P_{\text{sell}}^{[n|k]}
+ C_{\text{buy}}^{[n|k]} \big(P_{\text{buy}}^{[n|k]}+P_{\text{loss}}^{[n|k]}\big) \notag\\
&\quad\;\;+ C_{\text{buy}}^{[n|k]} \!\!\sum_{i\in\Omega_{\text{batt}}}\!\!
\big( {P}_{\text{ch,}i}^{[n|k]}(1{-}\eta_{\text{batt}}^{\text{ch}})
 + {P}_{\text{dis,}i}^{[n|k]}(1{-}\eta_{\text{batt}}^{\text{dis}}) \big) \notag\\
&\quad\;\;+ C_{\text{deg}}^{[n|k]} \!\!\sum_{i\in\Omega_{\text{batt}}}\!\!
\big({P}_{\text{ch,}i}^{[n|k]}+{P}_{\text{dis,}i}^{[n|k]}\big) + C_{\text{DG}}^{[n|k]}\!\sum_{i\in\Omega_{\text{DG}}}\!{P}_{\text{DG,}i}^{[n|k]}\Big)
\end{align}
which mainly considers the main grid selling revenue, purchasing cost, battery degradation cost, charging loss, and diesel generator cost over the prediction horizon. Additionally, the active power loss $P_{\mathrm{loss}}^{[n|k]}=\sum_{ij\in\Omega_{\mathrm{br}}}\tilde{l}_{ij}^{[n|k]} r_{ij}$ is treated as an extra purchase cost, thereby encouraging branch-loss reduction.
Details of the constraints at prediction step $n$ are listed below. 

\subsubsection{$C_{\mathrm{dyn}}$: Dynamic constraints}
These constraints are used to describe the trajectory of SoC, RES power (e.g., PV power), and load power in the prediction horizon.
\begin{subequations}
    \begin{align}
    &\sigma_{\text{SoC,}i}^{[n+1|k]}=\sigma_{\text{SoC,}i}^{[k|k]}+\sum_{d=k}^{n}\left(\frac{{P}_{\text{ch,}i}^{[d|k]}\,\eta_{\text{batt}}^{\text{ch}}}{E_{\text{max,}i}} - \frac{{P}_{\text{dis,}i}^{[d|k]}}{E_{\text{max,}i}\,\eta_{\text{batt}}^{\text{dis}}}\right)\Delta t_{\text{p}},\label{eq_SoC1}\\
    & {P}_{\text{res,}j}^{[n+1|k]}=f_{\text{R}}({P}_{\text{res,}j}^{[n|k]},P^{[n+1|k]}_{\text{dict,}i^*};\Delta t_{\text{p}}),\label{C4_3_0}\\
    &{P}_{\text{load,}j}^{[n+1|k]}={P}_{\text{load,}j}^{[n|k]}+f_{\text{L}}({P}_{\text{load,}j}^{[n|k]};\Delta t_{\text{p}}).\label{C4_3}
\end{align}
\end{subequations}
The SoC dynamic constraints in \eqref{eq_SoC1} is expressed in the summation form to make their linear dependence on decision variables explicit. Equations~\eqref{C4_3_0} and \eqref{C4_3} represent the forecasting models for renewable sources and loads, respectively. Power forecasting is performed before each MPC optimization iteration and incorporated as a fixed input sequence, thereby preserving the convexity of the resulting MPC problem. Details of the forecasting model are provided in Section~\ref{sec_forecastingmodel}.

\subsubsection{$C_{\text{alg}}$: Algebraic constraints}
These constraints are composed of the node injection power, power balance, and the power-flow model.
\begin{subequations}
    \begin{align}
    &{P}_{\text{batt,}j}^{[n|k]}={P}_{\text{dis,}j}^{[n|k]}-{P}_{\text{ch,}j}^{[n|k]},{P}_{\text{dis,}j}^{[n|k]}{P}_{\text{ch,}j}^{[n|k]}=0,\label{C4_1}\\
    & P_{\text{slk}}^{[n|k]}=P_{\text{buy}}^{[n|k]}-P_{\text{sell}}^{[n|k]},P_{\text{buy}}^{[n|k]}P_{\text{sell}}^{[n|k]}=0, \label{C4_2}\\
    & P_{j}^{[n|k]}
    = f_{\mathrm{grid}}(P_{ij}^{[n|k]},P^{[n|k]}_{jk',k'\in\mathcal{C}_{j}},I_{ij}^{[n|k]},r_{ij}) ,\label{C4_4}\\
    & \tilde{V}_{j}^{[n|k]}=f_{\mathrm{node}}(\tilde{V}_{i}^{[n|k]},I_{ij}^{[n|k]},r_{ij}),\label{C4_5}\\
    & \tilde{I}_{ij}^{[n|k]}=f_{\mathrm{branch}}(P_{ij}^{[n|k]},\tilde{V}_{i}^{[n|k]}).\label{C5}
\end{align}
\end{subequations}
where \eqref{C4_1} represents the battery power constraints and \eqref{C4_2} denotes the main grid power exchange constraints. These variable-decoupling strategies facilitate problem convexification. Constraint \eqref{C4_4} represents the power balance equation based on the detailed branch flow model, where $P_{j}=P_{\text{batt,}j}+P_{\text{res,}j}+P_{\text{load,}j}$ denotes the nodal power injection. Constraint \eqref{C4_5} corresponds to the nodal voltage equation, and \eqref{C5} describes the branch current equation. The constraints convexification will be addressed in Section~\ref{sec_BFmodel}. Note that all power variables are defined to be positive when injected into the grid, and their detailed definitions are provided in the Nomenclature section following the abstract.

\subsubsection{$C_{\text{ieq}}$: Inequality constraints}
These constraints consist of power-flow security constraints, state-of-charge (SoC) safety constraints, and SoC state preservation conditions.
\begin{subequations}
    \begin{align}
    & 0\le\tilde{l}_{ij}^{[n|k]}\le I_{ij,\max}^2,\quad V_{j,\min}^2\le\tilde{v}_{j}^{[n|k]}\le V_{j,\max}^2,\label{eq_boxConst1}\\
    & 0\leq P_{\text{buy}}^{[n|k]}\leq P_{\text{sys,max}},~0\leq P_{\text{sell}}^{[n|k]}\leq P_{\text{sys,max}},\label{eq_boxConst2}\\
    &0\leq{P}_{\text{dis,}j}^{[n|k]}\leq P_{\text{batt,max}},~0\leq{P}_{\text{ch,}j}^{[n|k]}\leq P_{\text{batt,max}},\label{eq_boxConst3}\\
    & P_{\text{DG,min}}\leq{P}_{\text{DG,}j}^{[n|k]}\leq P_{\text{DG,max}},\label{eq_boxConst4}\\
    & \sigma_{\text{min}}\leq\sigma_{\text{SoC,}i}^{[n|k]}\leq \sigma_{\text{max}},\label{eq_SoC2}\\
    &\sigma_{\text{SoC,}i}^{[0]}= \sigma_{\text{SoC,}i}^{[K_{\text{d}}]},\label{eq_SoC3}
    \end{align}
\end{subequations}
where $I_{ij,\max}$ is the current limit of the branch $ij$, $V_{j,\min}$ and $V_{j,\max}$ are the lower and upper bounds of the $j$th bus voltage, $P_{\text{sys,max}}$ is the power exchange limitation with the main grid, $P_{\text{batt,max}}$ is the battery power rating, and $P_{\text{DG,min}}$ combine with $P_{\text{DG,max}}$ are the boundaries of the DG power.
Constraint \eqref{eq_SoC2} bounds the SoC in a reasonable range.
Constraint \eqref{eq_SoC3} promotes day-to-day sustainability by preventing systematic under-charging across days. 

\section{Forecasting Model}\label{sec_forecastingmodel}

A kernel-based auto-regressive framework, i.e., kernel ridge regression (KRR), which utilizes the data efficiently and is suitable for a small sample scenario, is introduced first, and then the MMDG-AKRR is developed for solar power forecasting. 

\subsection{Kernel Ridge Regression}

The auto-regressive forecasting model, initialized at step $k$ within the MPC algorithm, is  
\begin{align}\label{eq_autoReg}
    P^{[n+1|k]}=P^{[n|k]}+f(P^{[n-N_{L}+1\to n|k]},t_n),
\end{align}
where $f$ maps the previous $N_{L}$-step power sequence, sampled at interval $\Delta t_{\text{p}}$, to the next-step increment $\Delta P$. In this study, $f$ is implemented using kernel ridge regression (KRR), shown in \eqref{eq_KRRseqForecast}, as it exhibits robust performance in small-sample settings and provides improved generalization, thereby reducing the risk of overfitting.
\begin{align}\label{eq_KRRseqForecast}
    f(\mathbf{x}_n)=\mathbf{k}^\top(\mathbf{x}_n)(\mathbf{K}+\lambda \mathbf{I})^{-1}\mathbf{y}_{\text{s}},
\end{align}
where $\mathbf{x}_n=[P^{[n-N_{L}+1\to n|k]},t_n]^\top\in\mathbb{R}^{N_L+1}$, $t_n$ is the normalized timestamp of the last step for the input power sequence $P^{[n-N_{L}+1\to n|k]}$, and $\lambda$ denotes the regularization coefficient. The label vector of samples is $\mathbf{y}_{\text{s}}=[\Delta P_{1},\dots,\Delta P_{N_{\text{sp}}}]^\top\in\mathbb{R}^{N_{\text{sp}}}$ with sample size $N_{\text{sp}}$. $\mathbf{K}\in\mathbb{R}^{N_{\text{sp}}\times N_{\text{sp}}}$ is the Gram matrix associated to the Gaussian kernel entries 
\begin{equation}
    k_{\text{ker}}(\mathbf{x}_i,\mathbf{x}_j)=\exp\!(-\frac{\|x_i - x_j\|_2^2}{2\sigma^2})
\end{equation}
with bandwidth $\sigma$. As an option, one can further incorporate multiple kernels to construct a linear-combination multi-kernel structure, known as multiple-KRR (MKRR)~\cite{MKRR_di2020}, which enhances model expressiveness theoretically at the cost of increased model complexity. $\mathbf{k}(\mathbf{x}_n)=[k_{\text{ker}}(\mathbf{x}_1,\mathbf{x}_n),\dots,k_{\text{ker}}(\mathbf{x}_{N_{\text{sp}}},\mathbf{x}_n)]^\top\in\mathbb{R}^{N_{\text{sp}}}$ represents the input kernel vector.

\begin{figure}
    \centering
    \includegraphics[width=0.7\linewidth]{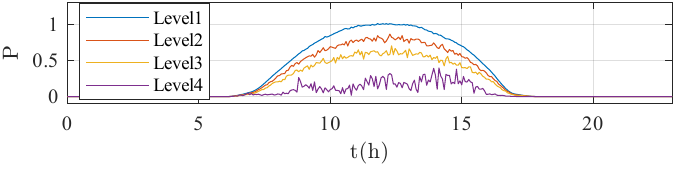}
    \caption{Demonstration of the diverse solar power profiles under varying weather conditions with different irradiation levels.}
    \label{fig_solarDiverseDemo}
\end{figure}

While the above KRR-based methods perform well in capturing regular patterns such as load profiles, renewable generation, particularly solar power, often exhibits highly diverse and multi-modal characteristics (Fig.~\ref{fig_solarDiverseDemo}). For highly variable PV generation profiles, forecasting models are typically trained using representative typical-day data that cover the most common operating conditions. As a result, such models often exhibit limited transferability to atypical or abrupt scenarios, which may lead to suboptimal planning decisions in the EMS. Moreover, the time-stamp feature differs from power-sequence features in both units and characteristic scales. Directly combining them as a single input may limit the expressive capability of KRR, because a kernel with a single bandwidth has limited flexibility in handling heterogeneous feature sensitivities. Thus, a more reliable forecasting method is required for the MPC-EMS over the prediction horizon. 

\subsection{Multi-Modal Dictionary-Guided Anisotropic-KRR}\label{sec_MMDGAKRR}

To address the above issues, this study proposes the MMDG-AKRR forecasting method. The main idea is to improve the representation of heterogeneous PV trajectories by combining multiple modal base models rather than fitting all scenarios with a single averaged forecasting model. Specifically, the nonzero time window of daily PV generation profiles is relatively consistent within the same region, and their overall shape exhibits several stable representative patterns. These characteristics motivate the construction of a base-model dictionary for typical groups of PV profiles, where each base AKRR model represents one typical PV operating modal, and the final prediction is obtained through online similarity-based weighting. 
{Since the prediction capability of a modal-dictionary-guided model depends on the representativeness of the predefined modals, the modal dictionary should cover the dominant PV trajectory patterns expected in the considered application. In general, the modal dictionary can be constructed either by clustering historical daily PV profiles, such as using K-means clustering, or by generating representative modal profiles from a typical-day template. The detailed dictionary-construction procedures are provided in~\ref{apdx_modal_dictionary}. }
It is noted that this approach intentionally neglects high-frequency details in the predicted profiles. However, such details are not critical for EMS applications that focus on power scheduling at relatively large time scales. 

The MMDG-AKRR forecasting model used at each prediction step of the MPC horizon, corresponding to \eqref{C4_3_0}, is formulated as
\begin{align}
P^{[n+1]} = P^{[n]} + \Xi^{[n]} \cdot \boldsymbol{\theta}(\mathbf{P}^{[n]}, t_n).
\label{eq_MMDGAKRR}
\end{align}
For clarity, the explicit $k$-step MPC notation adopted in \eqref{eq_autoReg} is omitted here. The weight vector associated with the $m$ representative PV profile modals is defined as $\Xi=[w_1,\dots,w_m]$, and the dictionary output vector is given by $\boldsymbol{\theta}(\mathbf{x}^{[n]}, t_n)=[f_1,\dots,f_m]^\top$. The input $\mathbf{P}^{[n]}=[P^{[n-N_{\mathrm{L}}+1\to n]}]$ denotes the initial power sequence of length $N_{\mathrm{L}}$.

Each component $f_i$ represents a base KRR model trained for the $i$th typical PV modal and equipped with an anisotropic Gaussian kernel defined as
\begin{align}
k_{\mathrm{ker}}\big([\mathbf{P}^{[i]},t_i],[\mathbf{P}^{[j]},t_j]\big)
=\exp\!\left(
-\frac{\|\mathbf{P}^{[i]}-\mathbf{P}^{[j]}\|_2^2}{2\sigma^2}
-\frac{\|t_i-t_j\|_2^2}{2\sigma_t^2}
\right),
\label{eq_AGK}
\end{align}
where $\sigma$ and $\sigma_t$ denote the bandwidth parameters associated with the power-sequence features and the time-stamp feature, respectively.

The AKRR hyperparameters $\lambda$, $\sigma$, and $\sigma_t$ are tuned using the proposed multi-start multi-step (MSMS) validation procedure, detailed in the~\ref{apdx_MSMS}. 

The overall procedure of the proposed MMDG-AKRR forecasting algorithm is summarized as follows.

\begin{figure*}[htbp]
\centering
\includegraphics[width=1\linewidth]{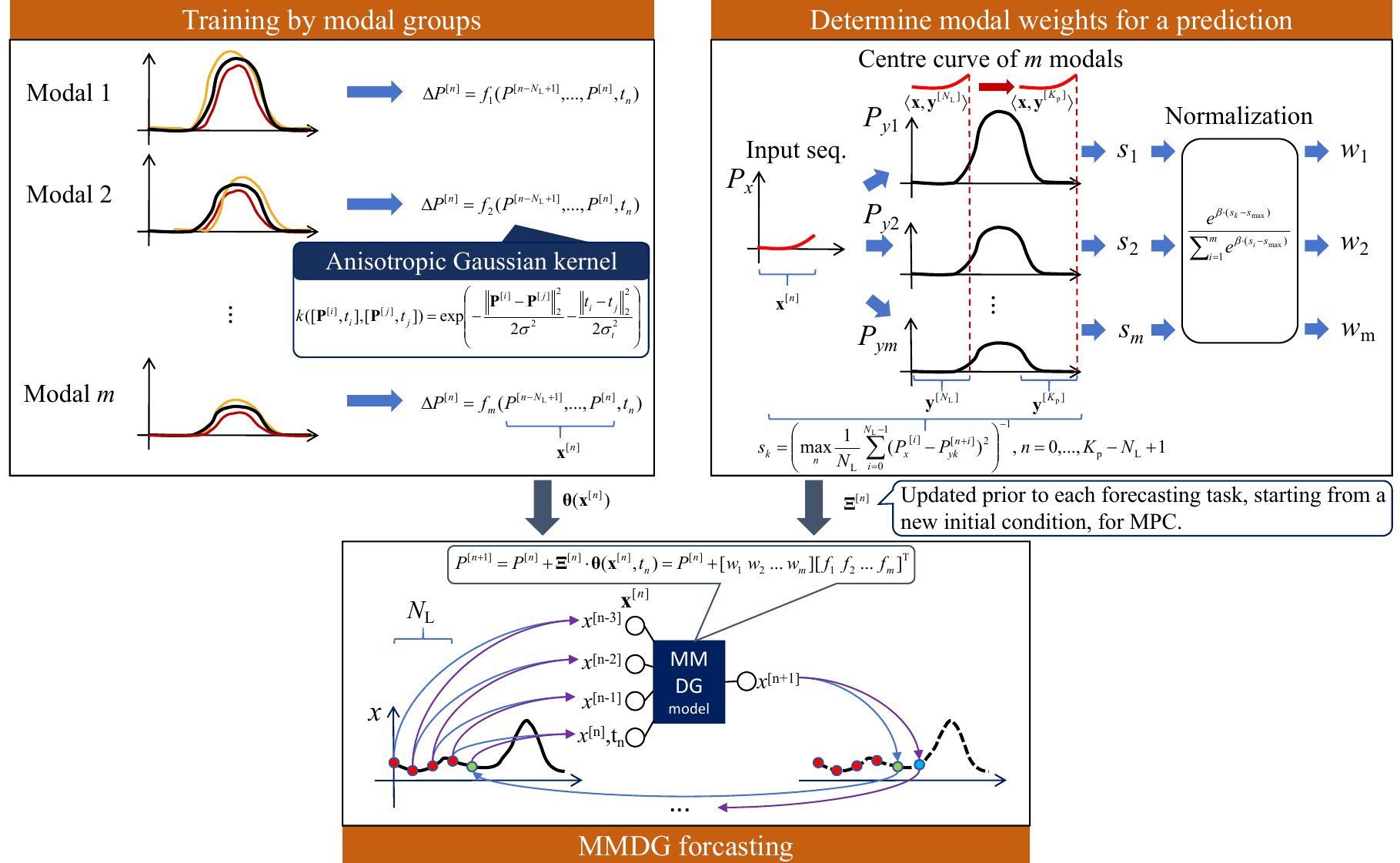}
\caption{Overall workflow of the proposed MMDG-AKRR forecasting algorithm.}
\label{fig_MMDGAKRR}
\end{figure*}

\begin{itemize}
\item \textbf{Step 1}: Classify the training power profiles into multiple typical modal groups, and decouple each profile into training sample pairs $\{[\mathbf{P}^{[i]},t_i],\Delta P_i\}_{i=1}^{N_{\text{sp}}}$.
    \item \textbf{Step 2}: Train a base anisotropic-KRR (AKRR) model $f$ for each modal group and assemble all base models into the dictionary $\boldsymbol{\theta}$.

\item \textbf{Step 3}: Prior to each prediction update within the MPC loop, evaluate the similarity between the online measured initial $N_{\text{L}}$-step input power sequence and the center profile of each training modal. The similarity between the initial sequence and the $k$th modal is defined as
\begin{align}
    s_k=\Bigg(\max_{n}\frac{1}{N_{\text{L}}}\sum_{i=0}^{N_{\text{L}}-1}
    \big(P_x^{[i]}-P_{y_k}^{[n+i]}\big)^2\Bigg)^{-1},
    \label{eq_similarity}
\end{align}
where $n=0,\dots,K_{\text{p}}-N_{\text{L}}+1$, $K_{\text{p}}$ denotes the prediction horizon length, $P_x^{[i]}$ is the $i$th element of the initial input sequence, and $P_{y_k}^{[n+i]}$ represents the $(n+i)$th point of the $k$th modal center profile. This metric evaluates the inverse of the minimum mean-square error (MSE) over all sliding segments, thereby identifying the most likely modal associated with the current initial condition. 
{
Note, the maximum similarity $S_{\max}=\max_k s_k$ can be used as an indicator of whether the current input sequence is sufficiently represented by the modal dictionary. A small $S_{\max}$ indicates that the sequence is far from all stored modal profiles, in which case the prediction should be interpreted as an out-of-dictionary forecast rather than a high-confidence modal match. In practice, a threshold for $S_{\max}$ can be selected from validation profiles or historical operating data to detect truly novel PV behaviors and trigger dictionary enrichment or online model updating when necessary.
}

\item \textbf{Step 4}: Normalize the similarity scores across all modals to obtain the weighting coefficients $w_k$ in $\Xi$ using
\begin{align}
    w_k=\frac{e^{\beta(s_k-s_{\max})}}{\sum_{i=1}^m e^{\beta(s_i-s_{\max})}},
    \label{eq_normalization}
\end{align}
where $s_{\max}=\max_i s_i$ and $\beta$ is a tuning parameter controlling the sharpness of the weight distribution. A larger $\beta$ enforces stronger concentration on the most similar modal. In this study, $\beta=0.01$ is adopted to achieve a sufficiently sharp distribution while maintaining numerical stability.

\item \textbf{Step 5}: Substitute the dictionary $\boldsymbol{\theta}$ and the weight vector $\Xi$ into \eqref{eq_MMDGAKRR}, and iteratively perform the forecasting update until the end of the prediction horizon.

\end{itemize}
The overall workflow of the proposed MMDG-AKRR algorithm is illustrated in Fig.~\ref{fig_MMDGAKRR}.

\section{Power-Flow-Constrained Convex MPC Formulation}\label{sec_BFmodel}


This section first introduces the simplified SOCP-relaxed power-flow/branch-flow model for the MPC-based EMS. Then, the standard conic-program form is organized.

\subsection{SOCP Relaxed Power-Flow Model}\label{sec_SOCP_relaxedmodel}

\begin{figure}
    \centering
    \includegraphics[width=0.35\linewidth]{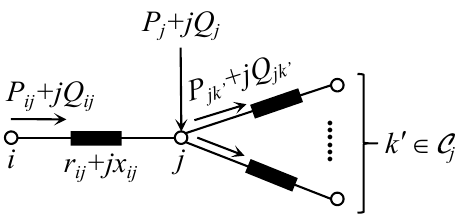}
    \caption{Notation of the branch flow model.}
    \label{fig_powerflowSketch}
\end{figure}

The power-flow model~\cite{Baran_Wu_IEEE_TPD1989,Farivar2013IEEE_TPS}, illustrated in Fig.~\ref{fig_powerflowSketch}, is given below
\begin{subequations}\label{eq_BKbranchpowerflow}
    \begin{align}
    P_{j} &= \sum_{k' \in \mathcal{C}_{j}} P_{jk'} - P_{ij} +  |\tilde{I}_{ij}| ^2 r_{ij}, \label{eq_BKbranchpowerflow1}\\[1ex]
    Q_{j} &= \sum_{k' \in \mathcal{C}_{j}} Q_{jk'} - Q_{ij} + |\tilde{I}_{ij}|^2 x_{ij}, \label{eq_BKbranchpowerflow2}\\[1ex]
    \tilde{V}_j &= \tilde{V}_i - \tilde{I}_{ij} \cdot (r_{ij} + j x_{ij}),\text{ with } \tilde{I}_{ij} =\frac{P_{ij} - j Q_{ij}}{\tilde{V}_i^*},\label{eq_BKbranchpowerflow3}
\end{align}
\end{subequations}

The non-convexity is introduced by $|\tilde{I}_{ij}|^2$ and $\tilde{I}_{ij}\tilde{V}_{i}^*$ terms.
Defining auxiliary variables $\tilde{v}=|\tilde{V}|^2$ and $\tilde{l}=|\tilde{I}|^2$, the relation $P_{ij}^2+Q_{ij}^2= \tilde{v}_i\tilde{l}_{ij}$ holds. To obtain a convex SOCP representation, the above equality is relaxed to $P_{ij}^2+Q_{ij}^2\leq \tilde{v}_i\tilde{l}_{ij}$, 
which can be rewritten as a second-order cone constraint $\|[2P_{ij},2Q_{ij},\tilde{v}_i-\tilde{l}_{ij}]\|_2\leq \tilde{v}_i+\tilde{l}_{ij}$.
To verify this, the following metric is proposed to quantify the relaxation gap,
\begin{equation}
    g_{t_{\mathrm d}}
    =
    \max_{(i,j)\in\Omega_{\mathrm{br}}}
    \frac{\big| P_{ij}(t_{\mathrm d})^2 - v_i(t_{\mathrm d}) \ell_{ij}(t_{\mathrm d}) \big|}
    {\max\!\left(P_{ij}(t_{\mathrm d})^2,\; v_i(t_{\mathrm d}) \ell_{ij}(t_{\mathrm d})\right)}
    \times 100\% ,
    \label{eq_relaxGap}
\end{equation}
where $g_{t_{\mathrm d}}$ denotes the maximum relative relaxation gap among all branches at decision step $t_{\mathrm d}$, reflecting the worst-case tightness of the SOCP relaxation.


In grid-connected operation, the main grid provides strong voltage and reactive power support, such that the microgrid EMS primarily focuses on active power scheduling for economic optimization. Under this condition, reactive power is typically regulated at lower control layers {and is not included as an EMS-level decision variable in this study}. Moreover, for distribution-level microgrids with relatively high $R/X$ ratios, i.e., $x_{ij} \ll r_{ij}$, the coupling between reactive power and system-wide economic objectives is further reduced in the grid-connected mode. {It should be noted that reactive power can still provide additional voltage-regulation capability. Therefore, neglecting reactive-power optimization may lead to a conservative assessment of the available voltage-regulation flexibility rather than an overestimation of voltage-security performance.} Thus, to reduce computational complexity while preserving sufficient accuracy for EMS-level {active-power scheduling}, the relaxed branch power-flow model in~\eqref{eq_BKbranchpowerflow} is simplified by neglecting reactive power optimization, leading to the following reduced formulation.

{If reactive-power dispatch or inverter volt-var control is required, the same MPC-EMS framework can be extended by retaining $Q_{ij}$ and the full $P$-$Q$ SOCP branch-flow constraints in \eqref{eq_BKbranchpowerflow}, at the expense of increased decision dimension and computational burden.}

\begin{subequations}\label{eq_SOCPsimp}
\begin{align}
    P_{j} = \sum_{k' \in \mathcal{C}_{j}} P_{jk'} - P_{ij} +  \tilde{l}_{ij} r_{ij}, \label{eq_SOCPsimp1}\\[1ex]
    \tilde{v}_j = \tilde{v}_i + \tilde{l}_{ij} \cdot r_{ij}^2-2P_{ij}r_{ij},\label{eq_SOCPsimp2}\\
    \|[2P_{ij},\tilde{v}_i-\tilde{l}_{ij}]\|_2\leq \tilde{v}_i+\tilde{l}_{ij},\label{eq_SOCPsimp3}
\end{align}
\end{subequations}
where \eqref{eq_SOCPsimp1}, \eqref{eq_SOCPsimp2}, and \eqref{eq_SOCPsimp3} are used to substitute the constraints \eqref{C4_4}, \eqref{C4_5}, and \eqref{C5} in the formulated problem.
Meanwhile, it provides the information for the active-power-loss term in the objective function, i.e., \(P_{\mathrm{loss}}^{[n|k]}=\sum_{ij\in\Omega_{\mathrm{br}}}\tilde{l}_{ij}^{[n|k]} r_{ij}\). 

\subsection{Conic Reformulation of MPC}\label{sec_STconicForm}

Several modeling details must be addressed before organizing the MPC problem into a standard conic form. First, to preserve convexity, the mutual exclusivity of the variable pairs $P_{\mathrm{dis}}$/$P_{\mathrm{ch}}$ and $P_{\mathrm{buy}}$/$P_{\mathrm{sell}}$ in \eqref{C4_1} and \eqref{C4_2} is not explicitly enforced. This is because the linear objective function naturally drives one variable in each pair to zero at optimality. 

Similarly, the terminal SoC equality constraint in \eqref{eq_SoC3} can be relaxed to the inequality $\sigma_{\text{SoC,}i}^{[0]}\leq \sigma_{\text{SoC,}i}^{[K_{\text{d}}]}$. This relaxation prevents systematic under-charging across scheduling horizons, {while the equality is typically tight under the positive TOU prices considered in this study because an unnecessarily high terminal SoC would require extra charging cost or reduced selling revenue. For special market conditions where such economic tightness may not hold, such as negative grid-purchase prices, the original terminal equality can be retained without changing the proposed MPC-EMS framework.}

After convexification, the resulting constraints are grouped into the following categories:
\begin{itemize}
    \item Box constraints  ($\mathbf{l}_{\text{bd}},\mathbf{u}_{\text{bd}}$): \eqref{eq_boxConst1}, \eqref{eq_boxConst2}, \eqref{eq_boxConst3}, \eqref{eq_boxConst4};
    \item SoC constraints ($\mathbf{A}_{\text{SoC}},\mathbf{b}_{\text{SoC}}$): \eqref{eq_SoC1}, \eqref{eq_SoC2}, \eqref{eq_SoC3}; 
    \item Power-flow constraints ($\mathbf{A}_{\text{grid}},\mathbf{b}_{\text{grid}}$): \eqref{C4_3_0}, \eqref{C4_3}, \eqref{C4_1}, \eqref{C4_2}, \eqref{eq_SOCPsimp1}, \eqref{eq_SOCPsimp2};
    \item Cone constraints ($\mathbf{A}_{\text{sc }i},\mathbf{b}_{\text{sc }i},\mathbf{d}_{\text{sc }i},\gamma_{\text{sc }i}$): \eqref{eq_SOCPsimp3}.
\end{itemize}

Aggregating the above constraints yields the following standard SOCP formulation of the MPC problem:
\begin{align}\label{eq_dayahead_SOCP_stform}
    &\min_{\mathbf{u}_{\mathrm{st}}}\; J=\mathbf{f}^\top\mathbf{u}_{\mathrm{st}}\\
    &\text{s.t.:}\,\notag\\
    &\mathbf{l}_{\text{bd}}\leq \mathbf{u}_{\mathrm{st}}\leq \mathbf{u}_{\text{bd}},\notag\\
    &\mathbf{A}_{\text{SoC}}\mathbf{u}_{\mathrm{st}}\leq\mathbf{b}_{\text{SoC}},\notag\\
    &\mathbf{A}_{\Delta E_{\sum}}\mathbf{u}_{\mathrm{st}}\leq\mathbf{b}_{\Delta E_{\sum}},\notag\\
    &\mathbf{A}_{\text{grid}}\mathbf{u}_{\mathrm{st}}=\mathbf{b}_{\text{grid}},\notag\\
    &\|\mathbf{A}_{\text{sc }i}\mathbf{u}_{\mathrm{st}}-\mathbf{b}_{\text{sc }i}\|_2\leq \mathbf{d}_{\text{sc }i}^\top \mathbf{u}_{\mathrm{st}}-\gamma_{\text{sc }i}, \, i=1,\dots,N_{\text{pre}} N_{\text{br}}\notag
\end{align}

{
\subsection{Discussion on SOCP Tightness under Weak-Grid and Islanded Conditions}

The SOCP branch-flow relaxation used in this paper follows the standard branch-flow relaxation framework, where the relaxation can be interpreted in two aspects: the conic relaxation of the squared voltage/current magnitude variables and the recovery of voltage/current phase angles. For the full $P$-$Q$ branch-flow model, the conic relaxation can still provide a convex and efficiently solvable formulation for the magnitude-related variables when the feasible region is nonempty. However, exact recovery of the voltage angles is topology-dependent. In particular, for radial distribution networks, the angle recovery is naturally satisfied under standard branch-flow relaxation assumptions, whereas for meshed networks, the angle condition must be explicitly checked or enforced through additional cycle constraints or convexification techniques.

The present study focuses on a grid-connected EMS setting, where the upstream grid provides the voltage reference and primary voltage support. Under islanded operation or weak-grid connection, this assumption becomes weaker or invalid, and inverter voltage/reactive-power support must be modeled more explicitly. Therefore, applying the proposed formulation to islanded or weak-grid microgrids does not necessarily invalidate the convex SOCP formulation itself, but it requires additional verification of angle recovery, relaxation exactness, and dynamic feasibility. In such cases, the full $P$-$Q$ branch-flow model, inverter voltage/reactive-power constraints, and post-optimization AC power-flow or dynamic validation should be included. This extension is beyond the grid-connected active-power EMS scope of this paper and is left for future work.
}

\section{Numerical Study}

This section first introduces the case setup. The effectiveness of the proposed MMDG-AKRR forecasting method is then evaluated. Finally, the proposed MMDG-AKRR-assisted MPC-EMS framework is validated in terms of operational security, cost reduction, and computational efficiency.

The proposed approach is evaluated on three microgrid systems. Detailed results are presented for the 18-bus system, while the results for the 10- and 33-bus systems are summarized subsequently.

\subsection{Case Setup}

\subsubsection{Load and Solar Profile}
\begin{figure}[htbp]
    \centering
    \subfloat[Solar power profile.]{%
        \includegraphics[width=0.4\linewidth]{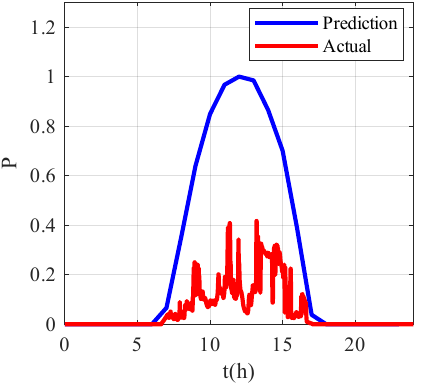}
        \label{fig_solarPower}
    }
    \hfil
    \subfloat[Load power profile for different types of users.]{%
        \includegraphics[width=0.4\linewidth]{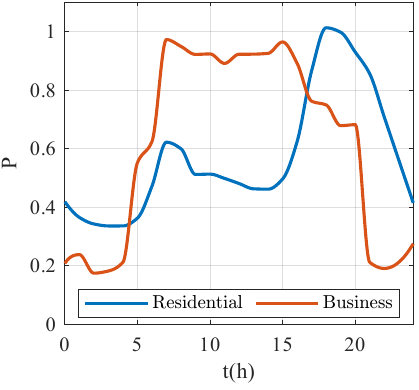}
        \label{fig_loadCurve}
    }
    \caption{Normalized power profiles used for the following tests.}
    \label{fig_solarLoadCurveDemo}
\end{figure}

To evaluate the effectiveness of the proposed MPC-based EMS under renewable generation uncertainty, the system performance is examined using a representative operating day characterized by pronounced solar forecast deviations, which commonly arise under variable irradiance conditions such as cloud coverage. This scenario is selected as an illustrative case to highlight the EMS behavior under challenging but realistic conditions, while similar performance trends are observed across other typical operating days. The normalized solar profile used for the following test is shown in Fig.~\ref{fig_solarLoadCurveDemo}(a), with data obtained from \cite{nie2023skipp}.
The normalized residential load and business load profiles used for the test are from \cite{OEDI_Dataset_153}, as shown in Fig. \ref{fig_solarLoadCurveDemo}(b). In the following cases, these normalized profiles are scaled to their respective rated power values.

\subsubsection{Battery Information}
The information on the battery used for the tests is as follows. The initial state of SoC is $\sigma_{\text{SoC}}^{[0]}=0.3$ with bounds $\sigma_{\text{max}}=0.9$ and $\sigma_{\text{min}}=0.2$. The charging and discharging efficiency are both $\eta_{\text{batt}}^{\text{ch}}=\eta_{\text{batt}}^{\text{dis}}=0.95$. The rated energy duration is $D_{\text{batt}}=4$h, and the life cycle is $N_{\text{cyc}}=5000$. The unit device cost is $C_{\text{unit}}=300\$$/kWh. The battery energy capacity is defined as $E_{\text{max}} = P_{\text{battR}} D_{\text{batt}}$, where $P_{\text{battR}}$ denotes the rated power. Specifically, $P_{\text{battR}}$ is set to 0.15~MW for all batteries in the 10- and 18-bus cases, while in the 33-bus case it is set to 0.2~MW for Battery~1 and 0.3~MW for Batteries~2--4.
 The degradation cost is $C_{\text{deg}}=0.5C_{\text{unit}}/N_{\text{cyc}}$.

\subsubsection{Electricity Price Information}
The electricity price information is provided in Table \ref{tab_price}.
\begin{table}[htbp]
    \caption{Price information.}
    \centering
    \small
    \begin{tabular}{M{2.5cm}|M{1.25cm}M{1.25cm}M{1.25cm}M{1.25cm}}
    \hline\hline
        Period & $C_{\text{buyRes}}$ (\$/kWh) & $C_{\text{buyBus}}$ (\$/kWh) & $C_{\text{sell}}$ (\$/kWh) & $C_{\text{DG}}$ (\$/kWh) \\
    \hline
        Valley & 0.10 & 0.06 & 0.02 & 0.30 \\
        Off-peak & 0.25 & 0.15 & 0.05 & 0.30 \\
        Peak & 0.35 & 0.25 & 0.10 & 0.30 \\
    \hline\hline
    \end{tabular}
    \label{tab_price}
\end{table}
Here $C_{\text{buyRes}}$ and $C_{\text{buyBus}}$ are the purchase tariffs for residential and business (commercial) loads, $C_{\text{sell}}$ is the sell price to the main grid, and $C_{\text{DG}}$ is the diesel generator cost.
The time-of-use (TOU) periods are defined as follows: Valley, 0h–8h; Off-peak, 8h–16h and 21h–24h; Peak, 16h–21h.

\subsubsection{Simulation Environment}

All simulations are conducted in MATLAB on a desktop computer equipped with an Intel Core i5-12400F CPU.
The validation problem in \eqref{eq_MSMS} for the kernel-based forecasting models is solved using a GA with a population size of 50 and a maximum of 30 generations, which provides a reasonable trade-off between solution quality and computational cost.
The EMS optimization problems are solved using either the \textit{linprog} or \textit{coneprog} solver, depending on the specific case configuration.
The MPC prediction horizon and the decision update interval are set to $\Delta t_{\mathrm{p}}=1 $ and $ \Delta t_{\mathrm{d}} = 0.5$ time units, which corresponds to one hour in this scenario, with a daily-length prediction horizon.
The simulation time step is set to $\Delta t = \Delta t_{\mathrm{p}}/900$.

\subsubsection{Comparison Methods}

The forecasting methods considered for performance comparison are summarized as follows:
\begin{itemize}
    \item \textbf{NN}: An auto-regressive forecasting framework in \eqref{eq_autoReg} equipped with a four-layer (128-64-32-1) neural-network (NN)-based inference model;
    \item \textbf{MMDG-NN}: The proposed MMDG framework equipped with the above neural-network-based inference model;
    \item \textbf{MKRR}: The proposed MMDG framework equipped with a conventional MKRR model using Gaussian and linear kernels as mentioned in \cite{DiWuMKRR2020};
    \item \textbf{AKRR}: The KRR forecasting model in \eqref{eq_KRRseqForecast} using an anisotropic Gaussian kernel;
    \item \textbf{MMDG-AKRR}: The proposed forecasting method described in Section~\ref{sec_MMDGAKRR};
\end{itemize}

The EMS frameworks considered for performance comparison are listed as follows:
\begin{itemize}
    \item \textbf{LP day-ahead EMS (LP-DA)}: A conventional linear-programming (LP) day-ahead EMS based on the AKRR forecasting model, which does not explicitly consider detailed branch-flow modeling and associated network security constraints;
    \item \textbf{SOCP day-ahead EMS (SOCP-DA)}: A power-flow-security-constrained day-ahead EMS based on the AKRR forecasting model, where the detailed branch-flow model is incorporated through an SOCP formulation;
    \item \textbf{NP day-ahead EMS (NP-DA)}: A power-flow-security-constrained day-ahead EMS based on the AKRR forecasting model, which adopts the original nonlinear power-flow model and is solved using an interior-point approach;
    \item \textbf{LP MPC-EMS (LP-MPC)}: An MPC version of the LP-based EMS;
    \item \textbf{SOCP MPC-EMS (SOCP-MPC)}: An MPC-based EMS equipped with the AKRR forecasting model and SOCP-relaxed power-flow constraints;
    \item \textbf{Proposed MPC-EMS (Proposed MPC)}: An MPC-based EMS equipped with the proposed MMDG-AKRR forecasting model and SOCP-relaxed power-flow constraints.
\end{itemize}

For all EMS evaluations, it is assumed that a reliable one-hour-ahead PV power forecast is available at each decision instant, enabled by real-time meteorological measurements/nowcasting and on-site irradiance sensing.

\subsection{Results of Forecasting Methods}

To quantify the forecasting accuracy, the mean absolute error (MAE) and the root mean square error (RMSE) are adopted to evaluate the deviation between the predicted power $\hat{P}$ and the reference trajectory $P_{\text{ref}}$, defined as
\begin{align}
    e_{\text{MAE}}=\frac{1}{K_{\text{p}}}\sum_{i=0}^{K_{\text{p}}}|\hat{P}^{[i]}-P^{[i]}_{\text{ref}}|,\,
    e_{\text{RMSE}}=\sqrt{\frac{1}{K_{\text{p}}}\sum_{i=0}^{K_{\text{p}}}(\hat{P}^{[i]}-P^{[i]}_{\text{ref}})^2}.\notag
\end{align}
In this study, multi-start forecasting is conducted by initializing the prediction at different time slots, i.e., $t_0 = 0, 5\!\sim\!17$~h. For each method, the overall forecasting performance is quantified by the average MAE and RMSE over all initial time instants, denoted as $\bar{e}_{\text{MAE}}$ and $\bar{e}_{\text{RMSE}}$, respectively.

{
To ensure generality, training samples are constructed using an auto-regressive sequence-to-point format based on the typical-day scaling strategy mentioned in~\ref{apdx_modal_dictionary}. Specifically, an input sample consists of the previous ($N_L$)-step PV power sequence and the normalized time stamp, while the output label is the next-step PV power increment. 
For the MMDG-based models, a modal dictionary is further constructed from scaled versions of the typical profile. In this study, the modal center profiles are generated as ($a_k P_{\mathrm{typ}}$), where ($a_k\in{1.0,0.9,\ldots,0.1}$). Around each modal center, three nearby profiles, ($0.96a_k P_{\mathrm{typ}}$), ($a_k P_{\mathrm{typ}}$), and ($1.04a_k P_{\mathrm{typ}}$), are used to form the training set for the corresponding base forecasting model. 
During online prediction, the initial measured PV sequence is compared with these modal center profiles, and the resulting similarity scores are used to determine the weights of the base models.
}

\subsubsection{Regular Day Forecasting}

\begin{figure*}[htbp]
    \centering
    \includegraphics[width=1\linewidth]{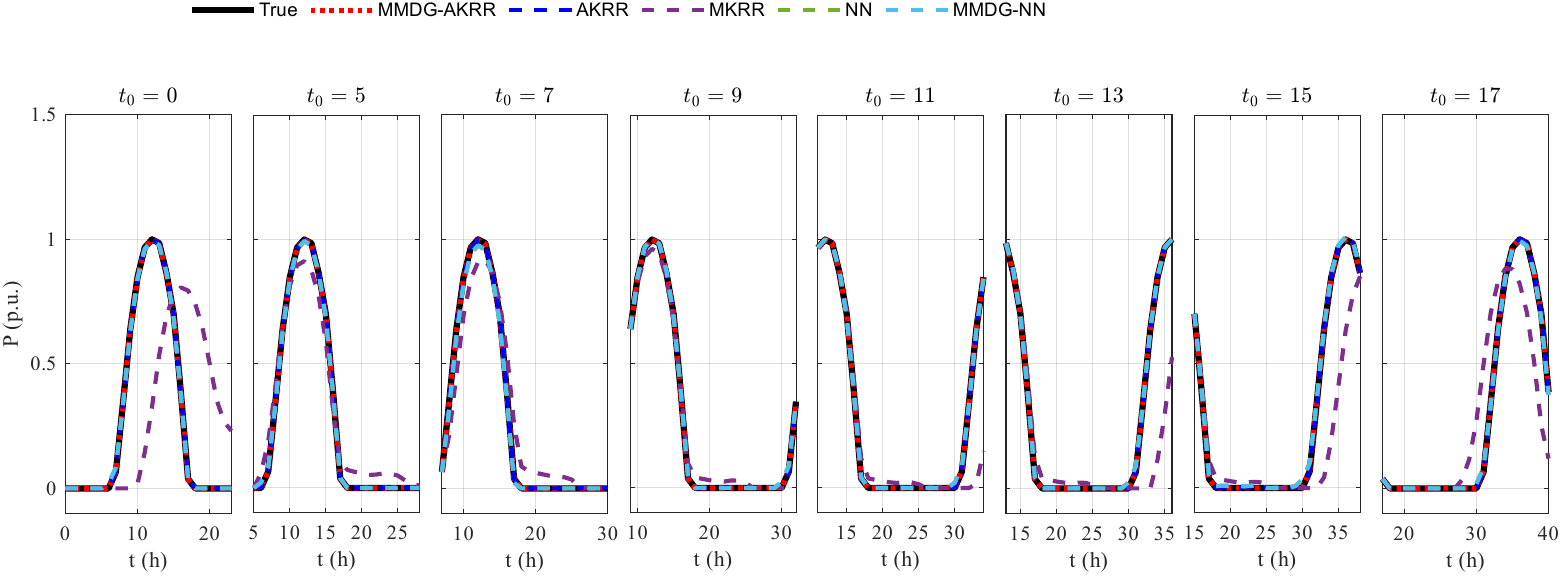}
    \caption{Shift-initial step forecasting results for a regular day.}
    \label{fig_multiStartForecasting_regular}
\end{figure*}

In this case, the test PV power profile is assumed to lie within the training data distribution. The multi-start forecasting results initialized at different time instants are illustrated in Fig.~\ref{fig_multiStartForecasting_regular}, and the corresponding averaged error metrics are summarized in Table~\ref{tab_multiStartForecasting_regular}.
\begin{table}[htbp]
    \caption{Forecasting Error For The Regular Day Case.}
    \centering
    \small
    \begin{tabular}{M{3cm}|M{1.5cm}M{1.5cm}}
    \hline\hline
        Methods & $\bar{e}_{\text{MAE}}$(\%) & $\bar{e}_{\text{RMSE}}$(\%)  \\
    \hline
    NN & 5.27e-1 & 9.18e-1  \\
    MMDG-NN & 5.27e-1 & 9.18e-1  \\
    MKRR & 10.52 & 18.03  \\
    AKRR   & 7.32e-2 & 1.07e-1  \\
    MMDG-AKRR & 7.32e-2 & 1.07e-1  \\
    \hline\hline
    \end{tabular}
    \label{tab_multiStartForecasting_regular}
\end{table}

As shown in Fig.~\ref{fig_multiStartForecasting_regular}, when the forecasting day is consistent with the training data distribution, all methods achieve relatively low prediction errors. Although MKRR exhibits higher error than the other approaches, it still captures the overall trend of the PV profile, indicating that the forecasting models are properly configured. The inferior performance of MKRR can be attributed to its single-bandwidth kernel, which provides limited flexibility in handling heterogeneous features such as the power sequence and time-stamp information, thereby highlighting the advantage of the anisotropic kernel adopted in the AKRR.

Moreover, based on Table~\ref{tab_multiStartForecasting_regular}, the MMDG-based methods exhibit identical performance to their corresponding baseline models. This is because, under regular operating conditions, the initial input sequence shows similar similarity to all modal profiles, leading the MMDG mechanism to assign a dominant weight to the representative typical-day model. As a result, the proposed MMDG framework remains fully compatible with conventional forecasting tasks under regular conditions.

\subsubsection{Irregular Day Forecasting}

\begin{figure*}[htbp]
    \centering
    \includegraphics[width=1\linewidth]{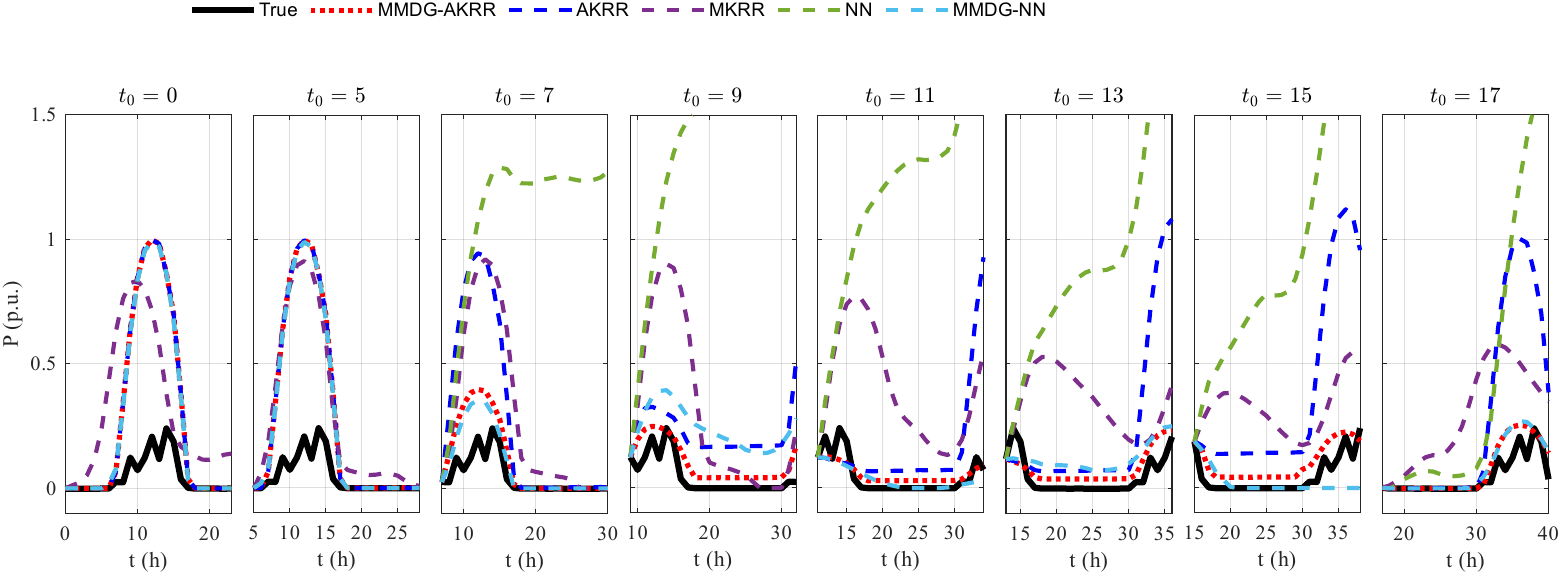}
    \caption{Shift-initial step forecasting results for an irregular day.}
    \label{fig_multiStartForecasting_irregular}
\end{figure*}

In contrast, this case considers an irregular PV power profile that lies outside the training data distribution, as illustrated in Fig.~\ref{fig_solarPower}. The corresponding multi-start forecasting results are presented in Fig.~\ref{fig_multiStartForecasting_irregular}, and the averaged error metrics are summarized in Table~\ref{tab_multiStartForecasting_irregular}.
\begin{table}[htbp]
    \caption{Forecasting Error For The Irregular Day Case.}
    \centering
    \small
    \begin{tabular}{M{3cm}|M{1.5cm}M{1.5cm}}
    \hline\hline
        Methods & $\bar{e}_{\text{MAE}}$(\%) & $\bar{e}_{\text{RMSE}}$(\%)  \\
    \hline
    NN & 82.59 & 96.13  \\
    MMDG-NN & 9.87 & 15.21  \\
    MKRR & 27.61 & 35.17  \\
    AKRR   & 22.22 & 34.16  \\
    MMDG-AKRR & 9.51 & 14.47  \\
    \hline\hline
    \end{tabular}
    \label{tab_multiStartForecasting_irregular}
\end{table}

As shown in Fig.~\ref{fig_multiStartForecasting_irregular}, when no reliable prior information is available for day-ahead forecasting, all methods initialize the prediction at $t_0 = 0$~h using models trained on typical modal profiles. Consequently, all approaches exhibit noticeable deviations from the true PV trajectory at the beginning of the day, particularly under abrupt and irregular operating conditions. As the MPC prediction horizon recedes and newly measured data are incorporated into the initial input sequence, the forecasting performance of different methods begins to diverge, reflecting their varying abilities to adapt to real-time changes.

Comparing the NN-based results in Tables~\ref{tab_multiStartForecasting_regular} and \ref{tab_multiStartForecasting_irregular}, it is evident that NN suffers a severe degradation in forecasting accuracy when the test profile deviates from the training distribution. Although AKRR also exhibits increased errors in the irregular-day scenario, its performance remains substantially better than that of NN, indicating stronger generalization capability and robustness to distribution shifts. Furthermore, both NN and AKRR benefit significantly from integration with the proposed MMDG framework, as reflected by the notable error reductions in Table~\ref{tab_multiStartForecasting_irregular}. This demonstrates that MMDG effectively enhances model adaptability by exploiting online updated information.

{
\subsubsection{Statistical Forecasting Evaluation on Multiple Irregular PV Profiles}
\label{sec_multi_irregular_forecasting}

The preceding irregular-day case provides a representative visualization of the forecasting behavior under a challenging PV trajectory. To further reduce the dependence on a single test profile, an additional statistical evaluation is conducted on multiple irregular PV profiles selected from the same dataset. These profiles cover different generation levels and shape deviations from the typical regular-day profile. 

To characterize the selected irregular PV profiles, three profile-level indicators are introduced. Let $P_{\mathrm{test}}(t)$ denote a selected irregular PV profile and $P_{\mathrm{typ}}(t)$ denote the typical regular-day profile used for constructing the modal dictionary. Both profiles are normalized by the peak value of the typical profile. The daily energy ratio is defined as
$EnergyRatio =
\frac{\sum_{t=1}^{24} P_{\mathrm{test}}(t)}
{\sum_{t=1}^{24} P_{\mathrm{typ}}(t)}$.
This indicator reflects the overall generation level of the selected irregular profile relative to the typical regular-day profile. The shape deviation is defined as
ShapeDeviation =
$\frac{1}{24}\sum_{t=1}^{24}
\left|
\frac{P_{\mathrm{test}}(t)}
{\max_t P_{\mathrm{test}}(t)+\epsilon}
\frac{P_{\mathrm{typ}}(t)}
{\max_t P_{\mathrm{typ}}(t)+\epsilon}
\right|$,
where $\epsilon$ is a small positive constant used to avoid division by zero. This indicator measures the profile-shape mismatch after removing the dominant amplitude effect. The ramping variation is defined as
$RampMean =
\frac{1}{23}\sum_{t=1}^{23}
\left|P_{\mathrm{test}}(t+1)-P_{\mathrm{test}}(t)\right|$,
which describes the average hourly fluctuation intensity of the PV trajectory. The characteristics of the selected irregular PV test profiles are summarized in Table~\ref{tab_irregular_pv_profile_features}.

\begin{table}[htbp]
\centering
\small
\caption{Characteristics of the selected irregular PV test profiles.}
\begin{tabular}{c|ccc}
\hline\hline
Profile & EnergyRatio (\%) & ShapeDeviation (\%) & RampMean (\%/h) \\
\hline
1 & 7.10 & 14.57 & 1.55 \\
2 & 17.66 & 7.83 & 2.18 \\
3 & 51.86 & 7.47 & 7.66 \\
4 & 60.45 & 7.02 & 7.96 \\
5 & 71.13 & 5.96 & 5.14 \\
6 & 34.51 & 12.18 & 5.30 \\
7 & 64.87 & 5.80 & 8.43 \\
8 & 27.57 & 9.49 & 4.67 \\
9 & 49.52 & 5.53 & 6.16 \\
10 & 83.22 & 2.70 & 7.87 \\
\hline\hline
\end{tabular}
\label{tab_irregular_pv_profile_features}
\end{table}

The statistical forecasting results are illustrated in Fig.~\ref{fig_multi_irregular_forecasting}. For each method, the bar height represents the mean value of the forecasting error over the selected irregular PV profiles. The lower and upper error bars indicate the deviations from the mean to the best-case and worst-case profile-wise errors, respectively, thereby reflecting the performance fluctuation across different irregular profiles. As shown in Fig.~\ref{fig_multi_irregular_forecasting}, MMDG-AKRR achieves the lowest averaged MAE and RMSE among all compared methods. Compared with AKRR, the proposed MMDG-AKRR further reduces both the mean forecasting error and the error fluctuation, indicating that the modal weighting mechanism improves the adaptability of AKRR under irregular PV trajectories. In addition, MMDG-NN also significantly improves the performance of the original NN baseline, although its accuracy remains comparable to the kernel-based baselines. These results suggest that the proposed modal-dictionary strategy is effective for reducing forecasting degradation when the tested PV trajectories deviate from the regular training profile.

\begin{figure}[htbp]
\centering
\begin{minipage}{0.48\linewidth}
\centering
\includegraphics[width=\linewidth]{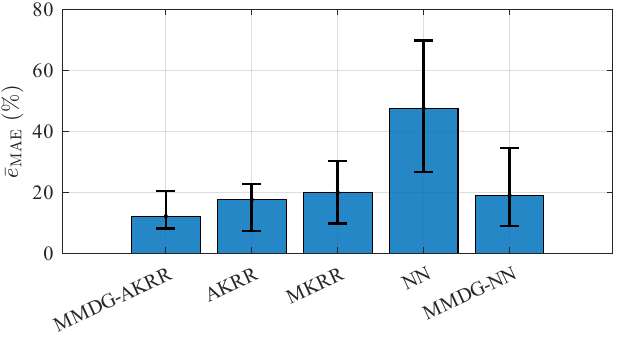}
\centerline{(a) MAE}
\end{minipage}
\hfill
\begin{minipage}{0.48\linewidth}
\centering
\includegraphics[width=\linewidth]{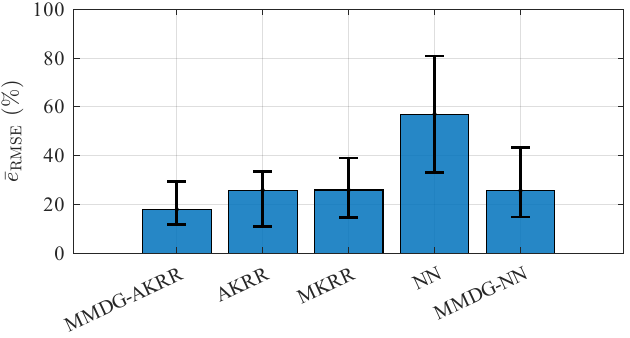}
\centerline{(b) RMSE}
\end{minipage}
\caption{Statistical forecasting performance on multiple irregular PV profiles.}
\label{fig_multi_irregular_forecasting}
\end{figure}

These results further verify that the advantage of the proposed MMDG-AKRR model is not limited to a single irregular PV trajectory. Therefore, the proposed forecasting model is adopted in the subsequent MPC-EMS study to provide improved online PV trajectory correction under forecast deviations.

}

Overall, MMDG-AKRR consistently achieves the lowest forecasting errors among all considered methods. Combined with the better generalization of AKRR over NN and conventional kernel-based approaches, these results support the adoption of MMDG-AKRR within the MPC-EMS framework.

\subsection{Results of EMS for the 18-Bus Microgrid}

\begin{figure}[htbp]
    \centering
    \includegraphics[width=0.7\linewidth]{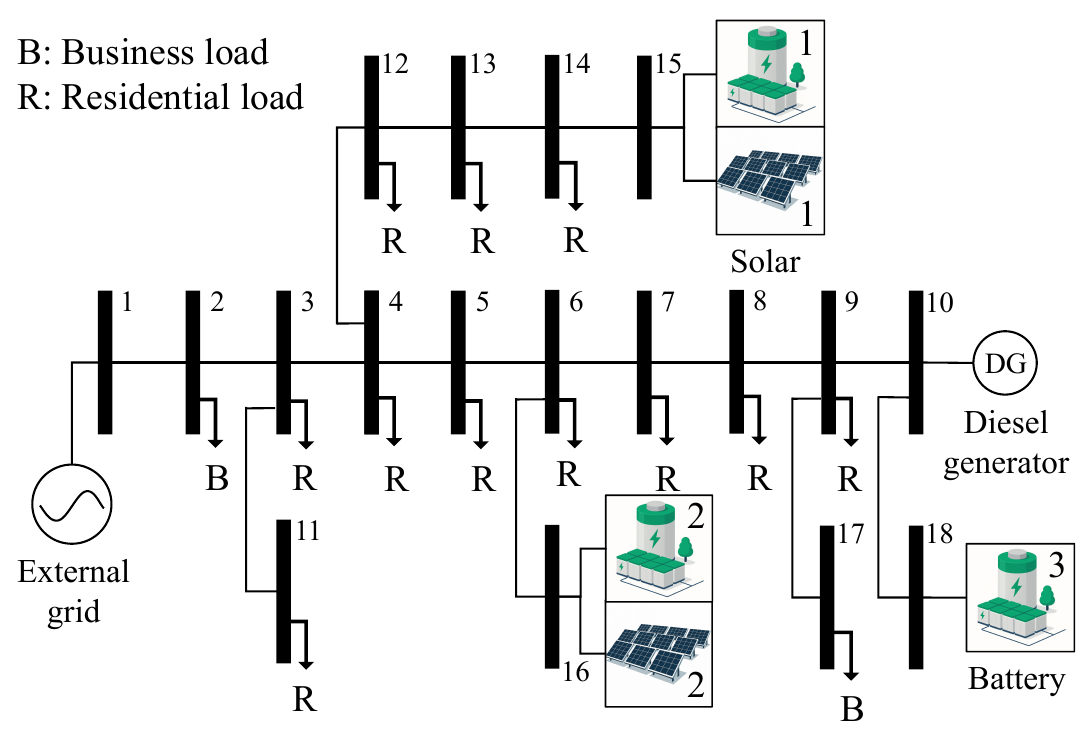}
    \caption{Topology of the CIRED 18-bus grid.}
    \label{fig_18bus_grid}
\end{figure}

This case is conducted on the CIGRE 18-bus low-voltage distribution benchmark microgrid system~\cite{CIGRE2014_Benchmark}, shown in Fig. \ref{fig_18bus_grid}.
The branch parameters are determined in accordance with~\cite {cerrowire_ampacity}. The power factor of the diesel generator is 0.85. The rated power for three batteries is 0.15MW. Solar panels connected to bus 15 and bus 16 have a rated power of 0.2MW and 0.3MW, respectively. To mitigate voltage drops caused by resistive losses along the long main feeder, a 0.1MW DG is installed at Bus~10, operating within a $20\%\text{–}100\%$ rated power range.
Aggregated residential users on each bus are assigned rated powers $P_{\mathrm{res}} = (10+20X)$~kW, where $X \sim \mathcal{U}(0,1)$. The rated power for the business users on buses 2 and 17 is 60 and 100kW, respectively.

\subsubsection{Grid safety results}

\begin{figure}[htbp]
    \centering
    \subfloat[Voltage profiles of the most under-voltage bus.]{%
        \includegraphics[width=0.45\linewidth]{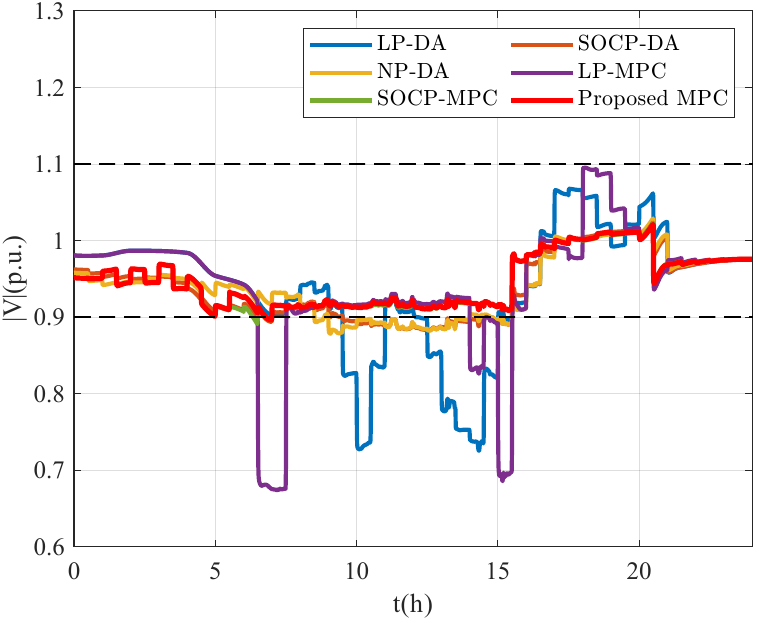}
        \label{fig_18busVoltageCurve}
    }
    \hfil
    \subfloat[Current profiles of the most overloaded branch.]{%
        \includegraphics[width=0.45\linewidth]{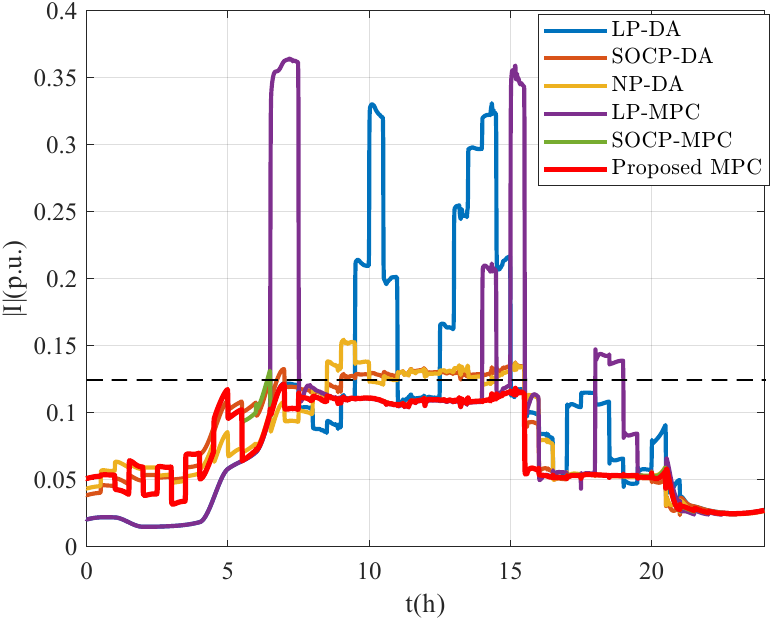}
        \label{fig_18busCurrentCurve}
    }
    \caption{Profiles of voltage and current.}
    \label{fig_voltageCurrentProfile}
\end{figure}

The bus voltage profiles for all six EMS cases are illustrated in Fig.~\ref{fig_18busVoltageCurve}, where the dashed line indicates the voltage security limit. The day-ahead LP-based EMS exhibits pronounced under-voltage during midday, whereas the LP MPC-based EMS suffers severe under-voltage issues in the morning and late afternoon. In contrast, the EMS schemes incorporating power-flow security constraints substantially reduce both the duration and magnitude of voltage violations. 

The branch current profiles of the most heavily loaded branch are shown in Fig.~\ref{fig_18busCurrentCurve}, where the dashed line denotes the maximum allowable current magnitude. As observed, both the LP day-ahead and LP MPC-based EMS cases exhibit large and sustained overcurrent violations, whereas methods constrained by power-flow security significantly suppress the current overlimit. In particular, the proposed MPC only shows a very small and short-duration current excursion under the strict pointwise threshold. Such minor post-simulation excursions are mainly caused by inverter tracking dynamics, interpolation, and finite simulation-step effects after the optimized dispatch is applied to the dynamic microgrid model, and they can be further mitigated by using a denser decision/simulation step in deployment. Overall, these results demonstrate the necessity of incorporating detailed grid-security constraints into EMS design.

{The detailed quantitative overlimit statistics are further reported in Appendix~\ref{apdx_grid_safety_quantification}.}


\subsubsection{Operation results}

\begin{figure}[htbp]
    \centering
    \includegraphics[width=1\linewidth]{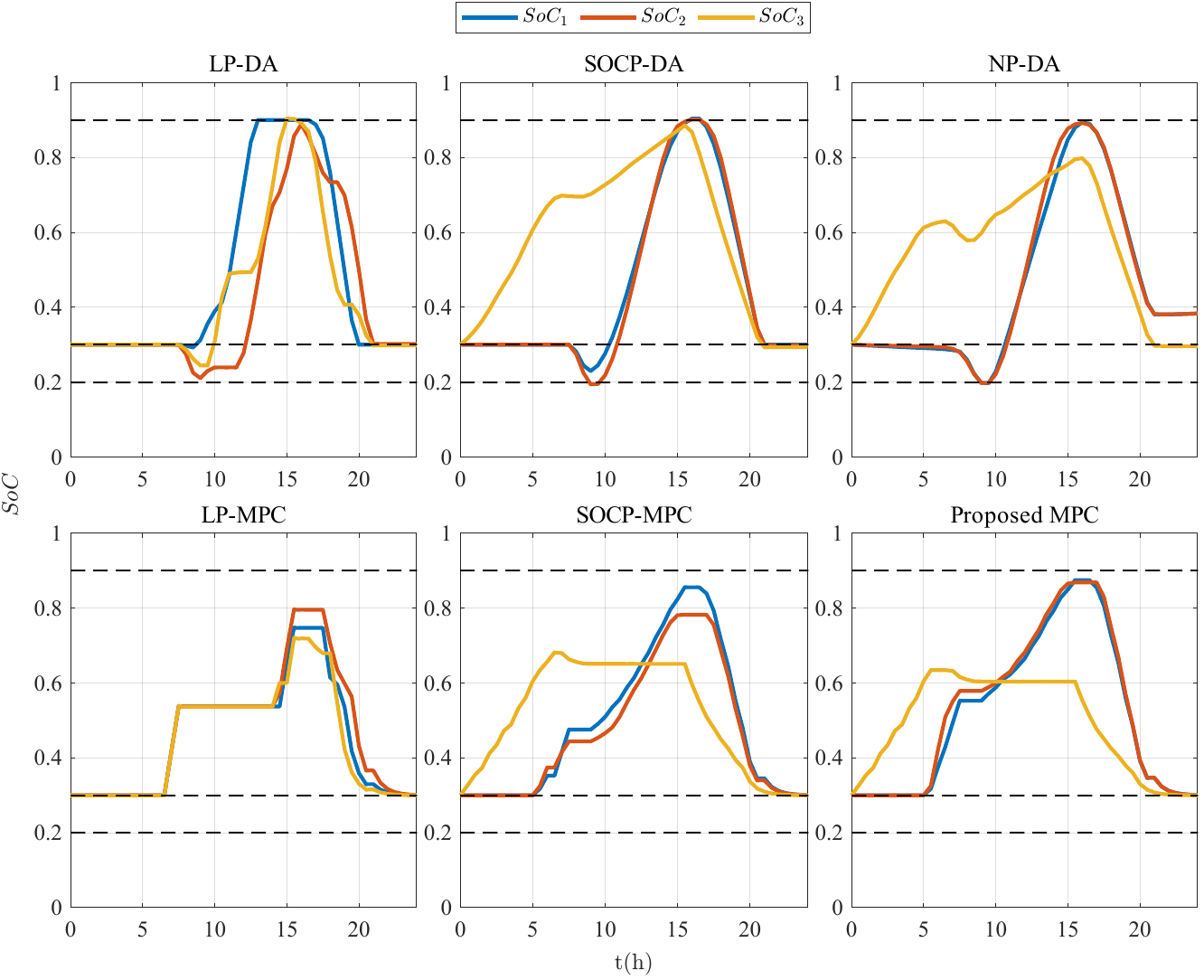}
    \caption{SoC profiles of batteries.}
    \label{fig_18busSoCCurve}
\end{figure}

The battery SoC trajectories for all considered cases are illustrated in Fig.~\ref{fig_18busSoCCurve}. As shown, the SoC of each battery remains within the prescribed bounds of $[0.2,0.9]$. For the convex EMS cases, the batteries are recharged close to their initial SoC levels at the end of the day, which is consistent with the relaxed terminal SoC requirement that the final SoC should be no lower than the initial SoC. {The only exception is the NP-DA benchmark. This abnormal terminal-SoC behavior should not be interpreted as looseness of the terminal-SoC relaxation at the optimum. Instead, NP-DA directly embeds the original nonlinear power-flow constraints into the EMS problem, which destroys convexity and makes the interior-point solver unable to guarantee global optimality under practical solver tolerances. Therefore, the NP-DA result reflects the numerical difficulty of directly solving the nonlinear power-flow-constrained EMS problem and further motivates the proposed convex SOCP-based formulation.} Compared with the two LP-based cases, the remaining methods charge the batteries more gradually over an extended time interval, thereby mitigating potential grid security issues. Moreover, for the three MPC-based EMS cases, the charging strategy is adaptively rescheduled in response to solar power shortages, avoiding unnecessary full charging of all batteries during non-valley periods and thus reducing additional operational costs.

\begin{figure}[htbp]
    \centering
    \includegraphics[width=1\linewidth]{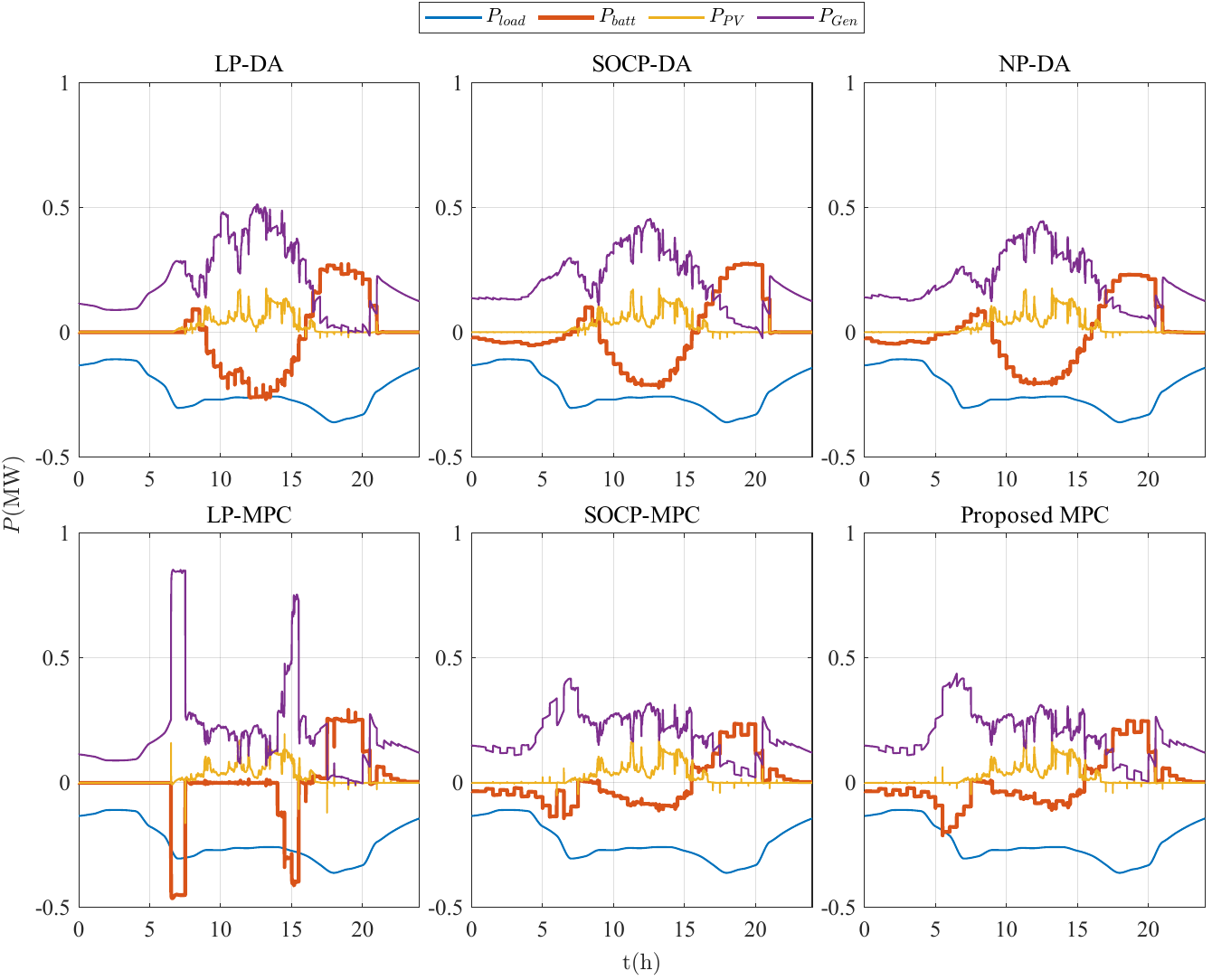}
    \caption{Aggregate power profiles for the four cases.}
    \label{fig_18busPowerCurve}
\end{figure}

Aggregated power profiles for each microgrid component are illustrated in Fig.~\ref{fig_18busPowerCurve}. All three day-ahead EMS schemes, namely LP-DA, SOCP-DA, and NP-DA, closely follow the day-ahead scheduling results. This leads to pronounced battery charging at noon (orange curves), which results in suboptimal operating costs, as discussed previously. For the LP-DA case, where power-flow constraints are not considered, such aggressive midday charging further induces branch overcurrent and bus under-voltage issues, as evidenced in Fig.~\ref{fig_18busVoltageCurve} and Fig.~\ref{fig_18busCurrentCurve}.

In contrast, the three MPC-based EMS methods—LP-MPC, SOCP-MPC, and the proposed MPC—can adaptively reschedule the battery charging strategy in response to real-time variations in solar power. In particular, these MPC schemes proactively increase battery charging during the valley-price period before anticipated solar power shortages, thereby reducing overall operating costs. However, without explicit power-flow constraints, the LP-MPC scheme still exhibits overly aggressive pre-noon charging behavior, which again leads to grid security violations. By comparison, the two power-flow-constrained MPC schemes distribute the charging power more smoothly over time, ensuring secure operation.

Furthermore, when comparing SOCP-MPC with the proposed MPC, the latter benefits from the improved forecasting accuracy provided by the MMDG-based prediction model. This enables the proposed MPC to more effectively leverage online information, shifting a greater portion of battery charging from the non-valley noon period to the early morning hours, thereby further reducing operating costs while maintaining grid security.

\subsubsection{Operation economic and time cost analysis}

Table~\ref{tab_18busCostResults} summarizes both the operating cost and the average decision-time cost per iteration, accounting for the entire prediction horizon.
By comparing the day-ahead EMS and the corresponding MPC-based EMS schemes (i.e., LP-DA vs. LP-MPC and SOCP-DA vs. SOCP-MPC), it can be observed that the MPC-EMS methods consistently achieve lower operating costs, since they can reschedule control actions online using updated system information.

Comparing power-flow-constrained methods with their non-constrained counterparts (i.e., LP-DA vs. SOCP-DA and LP-MPC vs. SOCP-MPC), the power-flow-constrained approaches yield lower costs. This improvement arises because EMS schemes with explicit power-flow models can exploit branch-level loss information and thereby enable more efficient power-flow optimization.

The proposed MPC-EMS method further reduces the operating cost by incorporating a forecasting model with higher sensitivity to modal variations during online updates, which leads to more accurate power predictions and improved scheduling decisions.

From a computational perspective, LP-based methods (LP-DA and LP-MPC) exhibit the highest time efficiency. The convex model-based methods, including SOCP-DA, SOCP-MPC, and the proposed MPC approach, also maintain relatively low computational costs and remain suitable for real-time implementation. In contrast, NP-DA, which relies on a nonlinear power-flow model, incurs a significantly higher computational burden and may suffer from suboptimal solutions. This comparison highlights the necessity and effectiveness of cone-relaxed power-flow constraints in achieving a favorable balance between optimality and computational efficiency.

Overall, the proposed MMDG-AKRR-assisted MPC-EMS method achieves the best overall performance by effectively reducing operating costs while ensuring grid security and maintaining time-efficient operation.
\begin{table}[htbp]
    \caption{Results for the 18-Bus Case.}
    \centering
    \small
    \begin{tabular}{M{2.5cm}|M{2cm}M{2cm}M{2cm}}
    \hline\hline
        Method & Cost($\$$) & Time/step(s) & PF const.  \\
         \hline
       LP-DA & 1122.62 & \textbf{8.95e-3} & Violated \\
       SOCP-DA & 1061.45 & \textbf{2.93e-1} & Critical \\
       NP-DA & 1106.60 & 943.86 & Critical \\
       LP-MPC & 1069.73 & \textbf{9.52e-3} & Violated \\
       SOCP-MPC & 1047.83 & \textbf{3.67e-1} & \textbf{Acceptable} \\
       Pro. MPC & \textbf{1030.32} & \textbf{3.11e-1} & \textbf{Acceptable} \\
         \hline\hline
    \end{tabular}
    \begin{minipage}{0.9\linewidth}
    \vspace{2pt}
    \footnotesize\emph{Note:} “PF const.” is an abbreviation for power-flow constraints, and "Pro. MPC" is an abbreviation for Proposed MPC.
    \end{minipage}
    \label{tab_18busCostResults}
\end{table}

{

\subsubsection{Prediction-Horizon Sensitivity Analysis}
\label{sec_prediction_horizon_sensitivity}

The performance of MPC-based EMS may depend on the selected prediction horizon because the controller optimizes the current action based on the predicted future PV generation, load demand, and electricity price. To examine the influence of this factor, an additional prediction-horizon sensitivity test is conducted. Three representative horizons are considered, namely 6 h, 12 h, and 24 h, corresponding to short-, medium-, and long-horizon settings, respectively. The decision update interval remains unchanged, and the controller is rolled forward over the complete 24 h operation period.

The baseline SOCP-MPC and the proposed MPC are compared under each prediction horizon. The total cost, average decision time, power-flow constraint status, and terminal SoC constraint status are summarized in Table~\ref{tab_prediction_horizon_ablation}. For the 6 h and 12 h horizons, the hard end-of-day terminal SoC constraint is not imposed inside every short receding window to avoid artificial infeasibility. Instead, the terminal SoC deficit after the complete 24 h operation is converted into an equivalent charging compensation penalty and included in the reported total cost. 
Specifically, the missing battery energy is compensated using the peak grid charging price while accounting for charging efficiency loss. 
Therefore, the reported cost provides a fair comparison even when the short- or medium-horizon case ends with a terminal SoC deficit. 
For the 24 h horizon, the original terminal SoC constraint is retained.

\begin{table}[htbp]
\centering
\small
\caption{Prediction-horizon sensitivity of the 18-bus MPC-EMS.}
\begin{tabular}{c|c|cccc}
\hline\hline
Horizon & Method & Cost (\$) & Time/step (s) & Terminal SoC status \\
\hline
6 h & SOCP-MPC & 1136.67 & 8.79e-2 & Violated \\ 
6 h & Pro. MPC & \textbf{1116.76} & 8.36e-2 & Violated \\ 
12 h & SOCP-MPC & 1104.12 & 1.78e-1 & Violated \\ 
12 h & Pro. MPC & \textbf{1082.16} & 1.78e-1 & Violated \\
24 h & SOCP-MPC & 1047.83 & 3.67e-1  & \textbf{Acceptable} \\
24 h & Pro. MPC & \textbf{1030.32} & 3.11e-1  & \textbf{Acceptable} \\
\hline\hline
\end{tabular}
\label{tab_prediction_horizon_ablation}
\end{table}

As shown in Table~\ref{tab_prediction_horizon_ablation}, the prediction horizon affects both the operating cost and computational burden. Shorter horizons reduce the optimization size and therefore require less decision time, but they provide less look-ahead information for future PV variation, load demand, and electricity-price changes. Meanwhile, shorter horizons also have difficulty in properly scheduling the batteries to meet the end-of-day terminal SoC target. Therefore, the terminal SoC deficits of the 6 h and 12 h cases are converted into equivalent charging compensation costs and included in the reported costs to avoid underestimating the cost of short-horizon operation. The 24 h horizon achieves the lowest cost because it can better coordinate battery charging and discharging over the whole daily cycle while satisfying the terminal SoC constraint. Across all tested horizons, the proposed MPC maintains a lower cost than the baseline SOCP-MPC, indicating that the advantage brought by the MMDG-AKRR-based PV correction is not limited to a specific prediction-horizon setting. In addition, the power-flow constraints remain satisfied in all tested cases, showing that the proposed online EMS preserves grid-security feasibility under different prediction-horizon configurations.

}

\subsection{Key Results for Other Microgrid Cases}

\begin{figure}[htbp]
    \centering
    \includegraphics[width=0.5\linewidth]{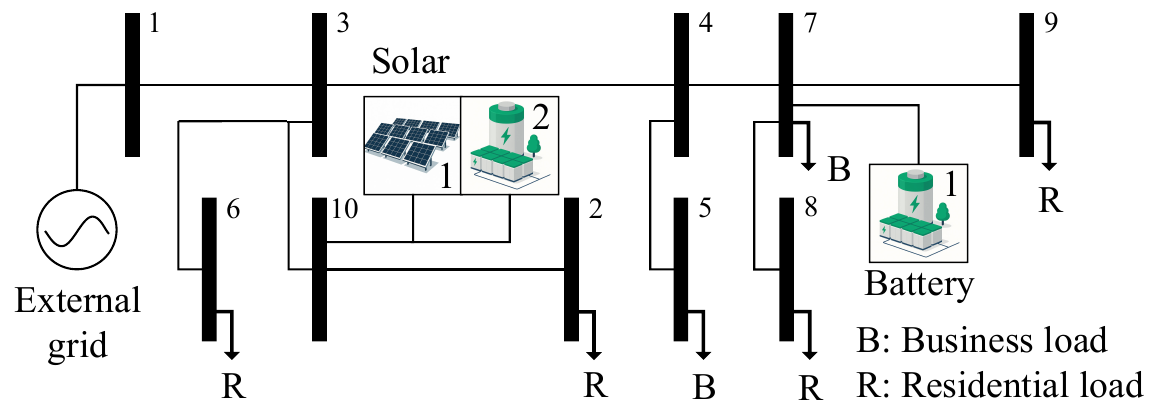}
    \caption{Topology of the 10-bus grid.}
    \label{fig_10bus_grid}
\end{figure}

\begin{figure}[htbp]
    \centering
    \includegraphics[width=0.8\linewidth]{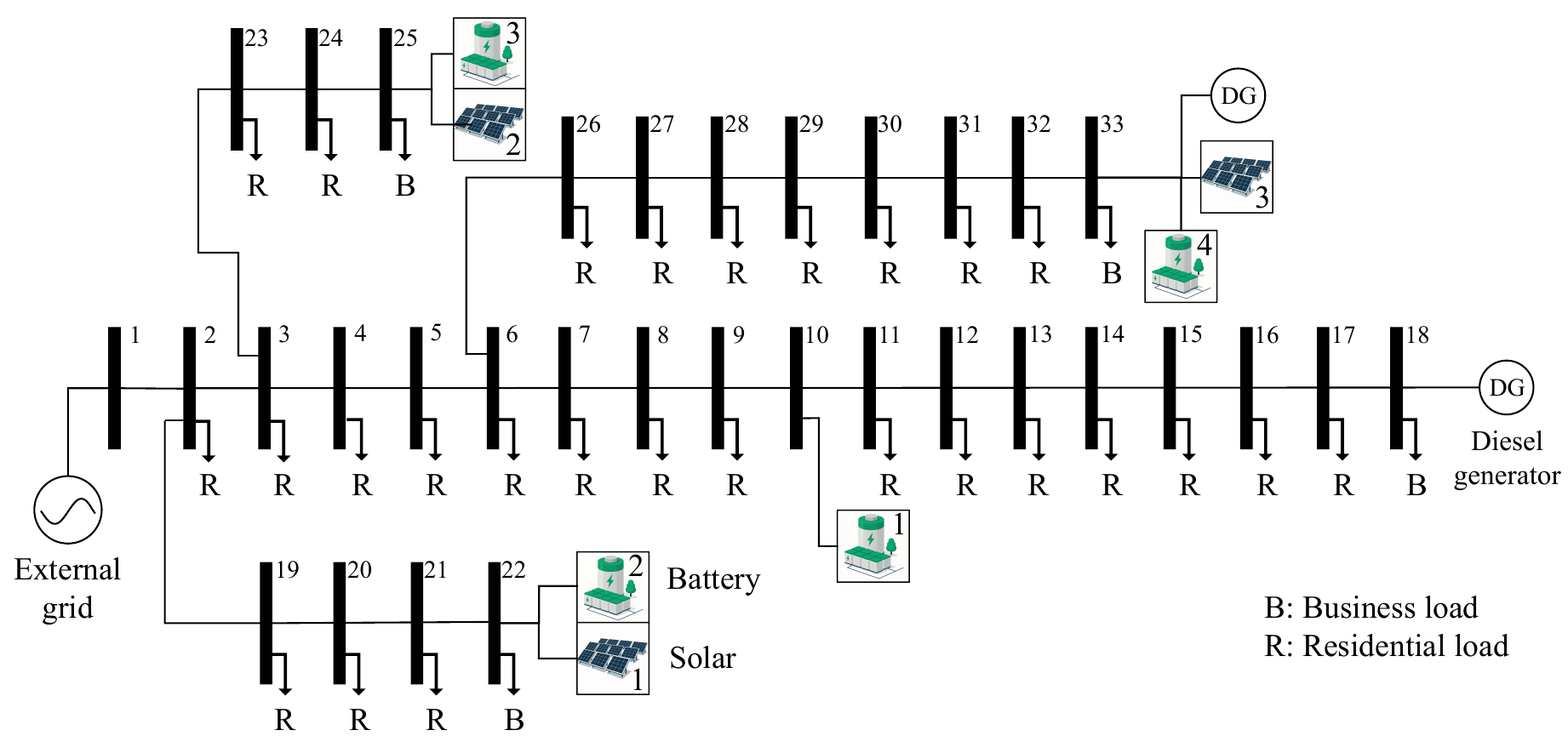}
    \caption{Topology of the IEEE 33-bus grid.}
    \label{fig_33bus_grid}
\end{figure}

To avoid over-specialization to a specific network configuration, the numerical tests described above are further conducted on {10-bus, 33-bus, and 69-bus distribution grids}, as shown in Fig.~\ref{fig_10bus_grid}, Fig.~\ref{fig_33bus_grid}, {and Fig.~\ref{fig_69bus_grid}}. 
{
\begin{figure}[htbp]
    \centering
    \includegraphics[width=0.8\linewidth]{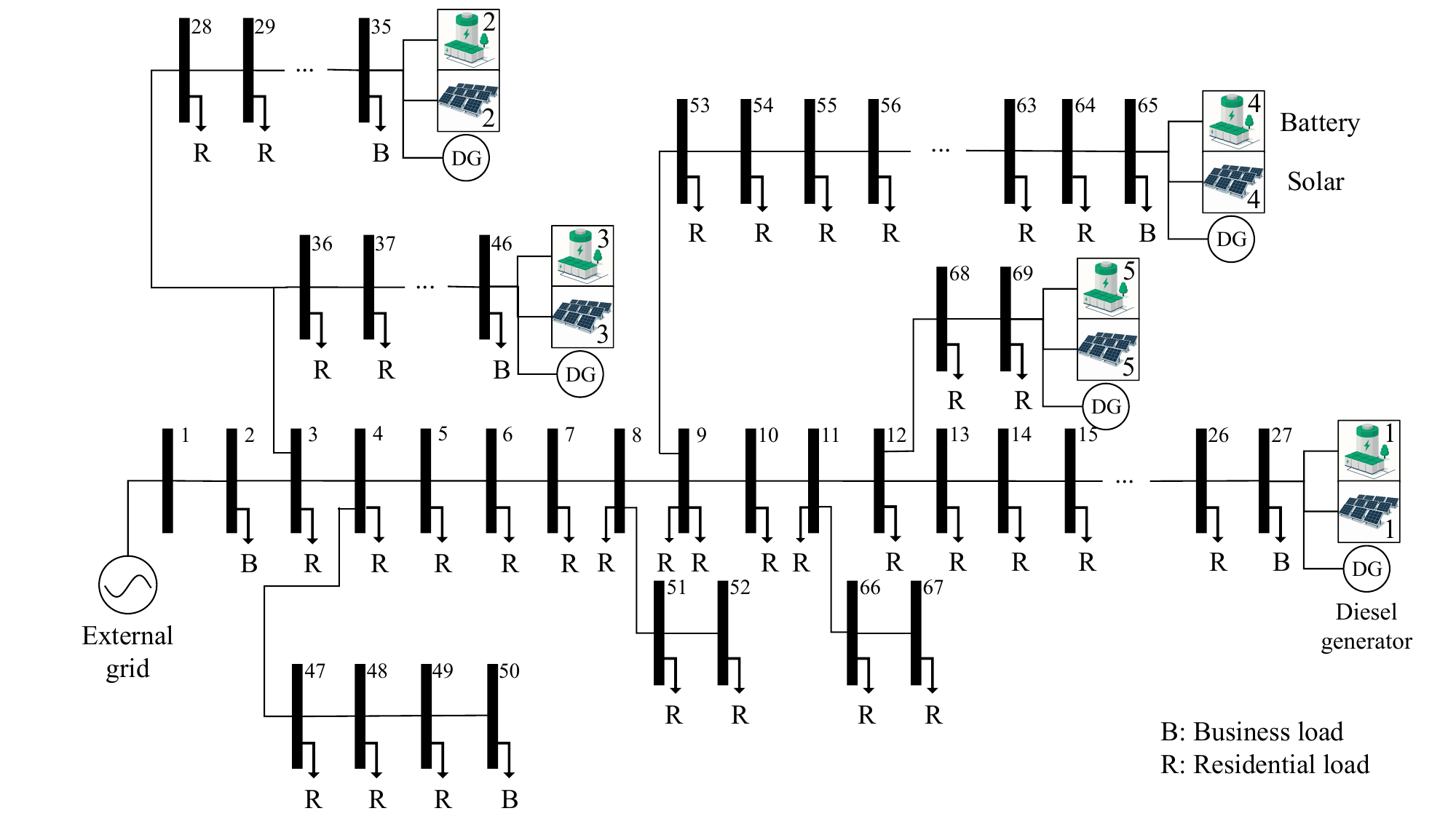}
    \caption{Topology of the IEEE 69-bus grid.}
    \label{fig_69bus_grid}
\end{figure}
}

The corresponding results are summarized in Table~\ref{tab_other_case_results}. Consistent conclusions can be drawn from these additional cases:
(1) MPC-based methods generally achieve better economic performance compared with day-ahead EMS approaches due to their enhanced adaptability to real-time uncertainties;
(2) Power-flow-constrained methods improve grid-security performance by explicitly accounting for branch-current and voltage constraints;
(3) Cone-relaxed power-flow formulations are essential for enabling the EMS optimization to efficiently obtain high-quality solutions, whereas the nonlinear power-flow-constrained formulation incurs a significantly higher computational burden; and
(4) The proposed MMDG-AKRR-assisted MPC-EMS delivers favorable overall performance by reducing operating costs while preserving grid-security performance and maintaining computational efficiency {across the tested benchmark systems}.

\begin{table}[htbp]
    \caption{{Results for the 10-Bus, 33-Bus, and 69-Bus Cases.}}
    \centering
    \small
    \setlength{\tabcolsep}{4pt}
    \begin{tabular}{M{1.4cm}M{2.4cm}|M{1.8cm}M{1.8cm}M{1.9cm}}
    \hline\hline
             Microgrid      &  Method & Cost($\$$) & Time/step(s) & PF const.  \\
         \hline
      \multirow{6}{*}{\raisebox{-1.5ex}{10-bus}} 
       & LP-DA & 464.03 & \textbf{4.78e-3} & Violated   \\
       & SOCP-DA & 457.13 & \textbf{2.07e-1} & \textbf{Acceptable}   \\
       & NP-DA & 481.39 & 268.24 & \textbf{Acceptable}   \\
       & LP-MPC & 452.56 & \textbf{1.10e-2} & Violated   \\
       & SOCP-MPC & 452.73 & \textbf{1.74e-1} & \textbf{Acceptable}   \\
       & Pro. MPC & \textbf{447.56} & \textbf{1.71e-1} & \textbf{Acceptable}   \\
        \hline
      \multirow{6}{*}{\raisebox{-1.5ex}{33-bus}} 
       & LP-DA & 2715.02 & \textbf{2.20e-2} &  Violated  \\
       & SOCP-DA & 2672.13 & \textbf{6.43e-1} & \textbf{Acceptable}   \\
       & NP-DA & 2819.71 & 1318.85 & \textbf{Acceptable}   \\
       & LP-MPC & 2776.58 & \textbf{1.26e-2} & Violated   \\
       & SOCP-MPC & 2676.11 & \textbf{7.20e-1} & \textbf{Acceptable}   \\
       & Pro. MPC & \textbf{2599.44} & \textbf{7.20e-1} & \textbf{Acceptable}   \\
        \hline
      \multirow{6}{*}{\raisebox{-1.5ex}{69-bus}} 
       & LP-DA & 4995.88 & \textbf{1.16e-2} & Violated   \\
       & SOCP-DA & 4980.60 & \textbf{9.56e-2} & Critical   \\
       & NP-DA & 5193.09 & 313.62 & Critical   \\
       & LP-MPC & \textbf{4822.69} & \textbf{1.13e-2} & Violated   \\
       & SOCP-MPC & 4933.45 & \textbf{2.27} & \textbf{Acceptable}   \\
       & Pro. MPC & \textbf{4829.01} & \textbf{2.48} & \textbf{Acceptable}   \\
         \hline\hline
    \end{tabular}
    \label{tab_other_case_results}
\end{table}


The tightness analysis for the cone-relaxation model can be found in{~\ref{apdx_cone_relaxation}.}

\section{Conclusion}


A multi-modal dictionary-guided anisotropic kernel ridge regression (MMDG-AKRR) forecasting model-based MPC-EMS is proposed to address solar-forecasting deviations caused by uncertain weather and associated power-flow-constraint violations. By combining multiple representative base models through an online similarity-driven weighting mechanism, the proposed forecasting framework {improves PV trajectory prediction under the represented multi-modal operating range}. To ensure grid-secure operation, cone-relaxed power-flow constraints are explicitly embedded into the MPC-EMS formulation, enabling reliable and computationally efficient scheduling decisions. Numerical results demonstrate that the proposed MMDG-AKRR model {achieves lower forecasting errors than the considered neural-network-based and kernel-based baselines under the tested regular, irregular, and multiple irregular PV profiles}, making it suitable for online MPC applications that rely on frequent forecast updates. Extensive case studies on 10-bus, 18-bus, 33-bus, {and 69-bus benchmark microgrids} verify that the proposed MMDG-AKRR-assisted MPC-EMS achieves {a favorable balance among operating cost, power-flow security, and computational efficiency}. These results highlight the {forecasting adaptability, operational effectiveness, and scalability-related computational tractability of the proposed framework across the tested benchmark systems} for microgrid energy management under high renewable uncertainty.

\appendix

{

\section{Modal Dictionary Construction for MMDG-AKRR}
\label{apdx_modal_dictionary}

The MMDG-AKRR framework requires a set of representative PV modal profiles to construct the base-model dictionary. In general, the modal dictionary can be obtained by either a data-driven clustering strategy or a typical-day scaling strategy.

For the clustering-based strategy, normalized daily PV profiles are grouped into $K$ clusters using profile-distance-based methods such as K-means clustering. The samples assigned to the $k$th cluster are used to train the corresponding base AKRR model, and the cluster-average profile is used as the modal center curve for online similarity evaluation. The number of modals $K$ determines the number of base models and is selected as a tradeoff between representation accuracy and computational complexity, which can be guided by validation performance.

For the typical-day scaling strategy, a representative clear-day profile $\mathbf{p}_{\mathrm{typ}}$ is first selected from historical PV profiles, for example by identifying a group of highly similar clear-day profiles and taking their average. Then, $K$ modal center profiles are generated by amplitude scaling:
\begin{align}
    \bar{\mathbf{P}}_k = a_k \mathbf{P}_{\mathrm{typ}}, \quad k=1,\dots,K,
\end{align}
where $a_k$ denotes the scaling factor used to cover different irradiance attenuation levels. Training samples around each scaled modal profile are then used to train the corresponding base AKRR model.

In the numerical study of this paper, the typical-day scaling strategy is adopted. This choice is suitable for the considered MPC-EMS application because PV generation has strong daily time-window regularity, and the equivalent PV energy over the prediction horizon is more influential for battery scheduling and grid-power exchange than local high-frequency profile fluctuations. Therefore, amplitude-scaled typical-day profiles provide a compact and practical modal dictionary for representing different PV generation levels in the tested EMS scenarios.

}

\section{Multi-Start Multi-Step Validation Procedure}\label{apdx_MSMS}

The objective function of the developed multi-start multi-step (MSMS) cross-validation is given in \eqref{eq_MSMS}. The multi-step property accounts for accumulated forecasting errors, while the multi-start property reduces sensitivity to initial points, which is essential for the MPC algorithm. 
\begin{align}\label{eq_MSMS}
    \min_{\sigma,\sigma_t,\lambda}\;\! \frac{1}{N_{v}}\!\sum_{v=1}^{N_v}\Big(\!\frac{1}{N_{s}}\!\sum_{s\in\Omega_s}\big(\!\frac{1}{N_{k}}\!\sum_{k=1}^{N_k} (\frac{x_k-(x_0+\sum_{i=1}^{k}f({\mathbf{x}}_n))}{x_k})^2\big)\!\Big)
\end{align}
where $\mathbf{x}_n$ denotes the input sequence, $x_0$ is the initial step, $x_k$ represents the $k$th step sample. $N_k$ is the number of forecasting steps. $N_s$ is the number of initial points, belonging to the set $\Omega_s$ and evenly distributed over a day. $N_v$ denotes the number of the validation fold. For each fold, the validation set is divided into a training fold and a test fold. The hyperparameters of $f$ are tuned on the training fold, while $x_k$ and $x_0$ are drawn from the test fold. The genetic algorithm (GA) is used as the solver for this validation process.

{

\section{Quantitative Grid-Safety Statistics}
\label{apdx_grid_safety_quantification}

To further quantify the grid-security performance shown in Fig.~\ref{fig_voltageCurrentProfile}, the overlimit duration and maximum violation magnitude of the displayed bus voltage and branch current are reported in Table~\ref{tab_appendix_grid_safety_stats}. The voltage violation is computed with respect to the lower and upper voltage limits, while the current violation is computed with respect to the allowable current magnitude of the displayed branch.

\begin{table}[htbp]
\centering
\small
\caption{Quantitative overlimit statistics of the displayed voltage and branch-current profiles.}
\begin{tabular}{c|cccc}
\hline\hline
Method 
& V-viol. time  
& Max. V viol. (p.u.) 
& I-viol. time  
& Max. I viol.  \\
\hline
LP-DA        & 19.47\% & 0.1745 & 16.74\% & 165.65\% \\
SOCP-DA      & 26.08\% & 0.0168 & 28.22\% & 10.42\% \\
NP-DA        & 25.88\% & 0.0217 & 24.55\% & 24.12\% \\
LP-MPC       & 10.35\% & 0.2262 & 12.64\% & 192.28\% \\
SOCP-MPC     & 1.55\%  & 0.0096 & 0.42\%  & 6.06\% \\
Pro. MPC     & 0.88\%  & 0.0035 & 0.00\%  & 0.00\% \\
\hline\hline
\end{tabular}
\label{tab_appendix_grid_safety_stats}
\end{table}

As shown in Table~\ref{tab_appendix_grid_safety_stats}, the LP-based EMS cases lead to relatively large voltage and current violations, especially in terms of the maximum violation magnitudes. The SOCP-constrained day-ahead EMS reduces the maximum voltage and current violation magnitudes compared with the LP-based day-ahead EMS, although it still exhibits non-negligible overlimit durations under the tested dynamic simulation. In comparison, the online SOCP-MPC cases further suppress both voltage and current violations. In particular, the proposed MPC achieves the smallest voltage-violation duration and magnitude, with only $0.88\%$ voltage-violation time and $0.0035$ p.u. maximum voltage violation. It also eliminates the displayed branch-current overlimit, with $0.00\%$ current-violation time and $0.00\%$ maximum relative current violation. These quantitative results confirm that the proposed MPC effectively suppresses sustained and severe grid-security violations under the tested scenario.

\section{Tightness Analysis of Cone-Relaxation}
\label{apdx_cone_relaxation}

The tightness of the SOCP relaxation is further examined under the rolling MPC framework. The obtained gap record forms a two-dimensional matrix, where each row corresponds to one rolling MPC decision and each column corresponds to one look-ahead step. This allows both the worst-case relaxation gap and the rolling stability of the relaxation tightness to be evaluated.

To avoid inflation of the relative gap caused by numerically inactive branches, branch records with near-zero relaxation terms are screened out. Specifically, a branch record is regarded as active only when its denominator is no smaller than $0.1\%$ of the maximum denominator among all branches at the same rolling decision and look-ahead step. 

The extended tightness statistics are summarized in Table~\ref{tab_appendix_relaxation_gap}. In addition to the mean and standard deviation, the table reports the 95th percentile, the maximum gap, and the time, look-ahead step, and branch location of the worst-case gap. The maximum first-step gap is also reported because only the first step of each MPC optimization is physically applied to the microgrid. To examine whether relaxation-tightening oscillations appear during rolling optimization, a first-step oscillation index is computed as the maximum adjacent variation of the first-step relaxation gap over consecutive MPC decisions.

\begin{table}[htbp]
\centering
\small
\setlength{\tabcolsep}{4pt}
\caption{Extended tightness statistics of the SOCP relaxation under rolling MPC.}
\begin{tabular}{l|cccc}
\hline\hline
Metric & 10-bus & 18-bus & 33-bus & 69-bus \\
\hline
Mean $\pm$ Std. (\%) & $1.52 \pm 1.01$ & $0.006 \pm 0.052$ & $0.007 \pm 0.044$ & $0.008 \pm 0.049$ \\
95th Perc. (\%) & 3.62 & 0.020 & 0.025 & 0.026 \\
Max. Gap (\%) & 3.98 & 4.07 & 3.49 & 2.46 \\
Worst Time (h) & 17.50 & 5.50 & 6.00 & 1.00 \\
Worst Look-ahead Step & 2 & 19 & 19 & 24 \\
Worst Branch & (7--8) & (3--4) & (26--27) & (1--2) \\
First-Step Max. (\%) & 3.96 & 0.236 & 0.222 & 0.122 \\
First-Step Osc. Index (\%) & 1.05 & 0.225 & 0.166 & 0.072 \\
\hline\hline
\end{tabular}
\label{tab_appendix_relaxation_gap}
\end{table}

As shown in Table~\ref{tab_appendix_relaxation_gap}, the relaxation gaps remain small for most operating points. The 95th-percentile gap is $3.62\%$, $0.020\%$, $0.025\%$, and $0.026\%$ for the 10-, 18-, 33-, and 69-bus cases, respectively. The maximum full-horizon gap is $3.98\%$, $4.07\%$, $3.49\%$, and $2.46\%$, respectively. For the 18-, 33-, and 69-bus cases, the worst full-horizon gaps occur at relatively far look-ahead steps, while the actually implemented first-step gaps remain much smaller. For the 10-bus case, the worst gap occurs at the second look-ahead step and is close to the maximum first-step gap, but its magnitude remains below $4\%$.

More importantly, the maximum first-step gap remains $3.96\%$, $0.236\%$, $0.222\%$, and $0.122\%$ for the four cases, respectively. Since only the first step of each rolling MPC solution is physically applied, these results indicate that the implemented dispatch decisions are only weakly affected by the cone relaxation. The first-step oscillation index is also limited to $1.05\%$, $0.225\%$, $0.166\%$, and $0.072\%$, respectively, showing that no evident relaxation-tightening oscillation appears during the rolling MPC process. These results support the satisfactory tightness of the SOCP relaxation for the proposed online EMS under the tested grid-connected scenarios, including the larger 69-bus system.

}



\bibliographystyle{elsarticle-num}  
\bibliography{Reference}

\end{document}